\documentclass[%
reprint,
nofootinbib,
amsmath,amssymb,
aps,
prl,
floatfix,
superscriptaddress
]{revtex4-2}

\usepackage{graphicx}
\usepackage{dcolumn}
\usepackage{bm}
\usepackage{hyperref}
\usepackage{comment}
\usepackage{xcolor}
\usepackage{subcaption}
\usepackage{siunitx}
\begin{document}


\preprint{APS/123-QED}

\title{Dual-Baseline Search for Active-to-Sterile Neutrino Oscillations in NOvA}

\newcommand{\ANL}{Argonne National Laboratory, Argonne, Illinois 60439, 
USA}
\newcommand{\Bandirma}{Bandirma Onyedi Eyl\"ul University, Faculty of 
Engineering and Natural Sciences, Engineering Sciences Department, 
10200, Bandirma, Balıkesir, Turkey}
\newcommand{\ICS}{Institute of Computer Science, The Czech 
Academy of Sciences, 
182 07 Prague, Czech Republic}
\newcommand{\IOP}{Institute of Physics, The Czech 
Academy of Sciences, 
182 21 Prague, Czech Republic}
\newcommand{\Atlantico}{Universidad del Atlantico,
Carrera 30 No.\ 8-49, Puerto Colombia, Atlantico, Colombia}
\newcommand{\BHU}{Department of Physics, Institute of Science, Banaras 
Hindu University, Varanasi, 221 005, India}
\newcommand{\UCLA}{Physics and Astronomy Department, UCLA, Box 951547, Los 
Angeles, California 90095-1547, USA}
\newcommand{\Caltech}{California Institute of 
Technology, Pasadena, California 91125, USA}
\newcommand{\Cochin}{Department of Physics, Cochin University
of Science and Technology, Kochi 682 022, India}
\newcommand{\Charles}
{Charles University, Faculty of Mathematics and Physics,
 Institute of Particle and Nuclear Physics, Prague, Czech Republic}
\newcommand{\Cincinnati}{Department of Physics, University of Cincinnati, 
Cincinnati, Ohio 45221, USA}
\newcommand{\CSU}{Department of Physics, Colorado 
State University, Fort Collins, CO 80523-1875, USA}
\newcommand{\CTU}{Czech Technical University in Prague,
Brehova 7, 115 19 Prague 1, Czech Republic}
\newcommand{\Dallas}{Physics Department, University of Texas at Dallas,
800 W. Campbell Rd. Richardson, Texas 75083-0688, USA}
\newcommand{\DallasU}{University of Dallas, 1845 E 
Northgate Drive, Irving, Texas 75062 USA}
\newcommand{\Delhi}{Department of Physics and Astrophysics, University of 
Delhi, Delhi 110007, India}
\newcommand{\JINR}{Joint Institute for Nuclear Research,  
Dubna, Moscow region 141980, Russia}
\newcommand{\Erciyes}{
Department of Physics, Erciyes University, Kayseri 38030, Turkey}
\newcommand{\FNAL}{Fermi National Accelerator Laboratory, Batavia, 
Illinois 60510, USA}
\newcommand{\FSU}{Florida State University, Tallahassee, Florida 32306, USA}
\newcommand{\UFG}{Instituto de F\'{i}sica, Universidade Federal de 
Goi\'{a}s, Goi\^{a}nia, Goi\'{a}s, 74690-900, Brazil}
\newcommand{\Guwahati}{Department of Physics, IIT Guwahati, Guwahati, 781 
039, India}
\newcommand{\Harvard}{Department of Physics, Harvard University, 
Cambridge, Massachusetts 02138, USA}
\newcommand{\Houston}{Department of Physics, 
University of Houston, Houston, Texas 77204, USA}
\newcommand{\IHyderabad}{Department of Physics, IIT Hyderabad, Hyderabad, 
502 205, India}
\newcommand{\Hyderabad}{School of Physics, University of Hyderabad, 
Hyderabad, 500 046, India}
\newcommand{\IIT}{Illinois Institute of Technology,
Chicago IL 60616, USA}
\newcommand{\Indiana}{Indiana University, Bloomington, Indiana 47405, 
USA}
\newcommand{\INR}{Institute for Nuclear Research of Russia, Academy of 
Sciences 7a, 60th October Anniversary prospect, Moscow 117312, Russia}
\newcommand{\Iowa}{Department of Physics and Astronomy, Iowa State 
University, Ames, Iowa 50011, USA}
\newcommand{\Irvine}{Department of Physics and Astronomy, 
University of California at Irvine, Irvine, California 92697, USA}
\newcommand{\Jammu}{Department of Physics and Electronics, University of 
Jammu, Jammu Tawi, 180 006, Jammu and Kashmir, India}
\newcommand{\Lebedev}{Nuclear Physics and Astrophysics Division, Lebedev 
Physical 
Institute, Leninsky Prospect 53, 119991 Moscow, Russia}
\newcommand{\Magdalena}{Universidad del Magdalena, Carrera 32 No 22-08 Santa Marta, Colombia}
\newcommand{\MSU}{Department of Physics and Astronomy, Michigan State 
University, East Lansing, Michigan 48824, USA}
\newcommand{\Crookston}{Math, Science and Technology Department, University 
of Minnesota Crookston, Crookston, Minnesota 56716, USA}
\newcommand{\Duluth}{Department of Physics and Astronomy, 
University of Minnesota Duluth, Duluth, Minnesota 55812, USA}
\newcommand{\Minnesota}{School of Physics and Astronomy, University of 
Minnesota Twin Cities, Minneapolis, Minnesota 55455, USA}
\newcommand{\Mississippi}{University of Mississippi, University, Mississippi 38677, USA}
\newcommand{\NISER}{National Institute of Science Education and Research,
Khurda, 752050, Odisha, India}
\newcommand{\Oxford}{Subdepartment of Particle Physics, 
University of Oxford, Oxford OX1 3RH, United Kingdom}
\newcommand{\Panjab}{Department of Physics, Panjab University, 
Chandigarh, 160 014, India}
\newcommand{\Pitt}{Department of Physics, 
University of Pittsburgh, Pittsburgh, Pennsylvania 15260, USA}
\newcommand{\QMU}{Particle Physics Research Centre, 
Department of Physics and Astronomy,
Queen Mary University of London,
London E1 4NS, United Kingdom}
\newcommand{\RAL}{Rutherford Appleton Laboratory, Science 
and 
Technology Facilities Council, Didcot, OX11 0QX, United Kingdom}
\newcommand{\SAlabama}{Department of Physics, University of 
South Alabama, Mobile, Alabama 36688, USA} 
\newcommand{\Carolina}{Department of Physics and Astronomy, University of 
South Carolina, Columbia, South Carolina 29208, USA}
\newcommand{\SDakota}{South Dakota School of Mines and Technology, Rapid 
City, South Dakota 57701, USA}
\newcommand{\SMU}{Department of Physics, Southern Methodist University, 
Dallas, Texas 75275, USA}
\newcommand{\Stanford}{Department of Physics, Stanford University, 
Stanford, California 94305, USA}
\newcommand{\Sussex}{Department of Physics and Astronomy, University of 
Sussex, Falmer, Brighton BN1 9QH, United Kingdom}
\newcommand{\Syracuse}{Department of Physics, Syracuse University,
Syracuse NY 13210, USA}
\newcommand{\Tennessee}{Department of Physics and Astronomy, 
University of Tennessee, Knoxville, Tennessee 37996, USA}
\newcommand{\Texas}{Department of Physics, University of Texas at Austin, 
Austin, Texas 78712, USA}
\newcommand{\Tufts}{Department of Physics and Astronomy, Tufts University, Medford, 
Massachusetts 02155, USA}
\newcommand{\UCL}{Physics and Astronomy Department, University College 
London, 
Gower Street, London WC1E 6BT, United Kingdom}
\newcommand{\Virginia}{Department of Physics, University of Virginia, 
Charlottesville, Virginia 22904, USA}
\newcommand{\WSU}{Department of Mathematics, Statistics, and Physics,
 Wichita State University, 
Wichita, Kansas 67260, USA}
\newcommand{\WandM}{Department of Physics, William \& Mary, 
Williamsburg, Virginia 23187, USA}
\newcommand{\Wisconsin}{Department of Physics, University of 
Wisconsin-Madison, Madison, Wisconsin 53706, USA}
\newcommand{\deceased}{Deceased.}
\affiliation{\ANL}
\affiliation{\Atlantico}
\affiliation{\Bandirma}
\affiliation{\BHU}
\affiliation{\Caltech}
\affiliation{\Charles}
\affiliation{\Cincinnati}
\affiliation{\Cochin}
\affiliation{\CSU}
\affiliation{\CTU}
\affiliation{\Delhi}
\affiliation{\Erciyes}
\affiliation{\FNAL}
\affiliation{\FSU}
\affiliation{\UFG}
\affiliation{\Guwahati}
\affiliation{\Houston}
\affiliation{\Hyderabad}
\affiliation{\IHyderabad}
\affiliation{\IIT}
\affiliation{\Indiana}
\affiliation{\ICS}
\affiliation{\INR}
\affiliation{\IOP}
\affiliation{\Iowa}
\affiliation{\Irvine}
\affiliation{\JINR}
\affiliation{\Magdalena}
\affiliation{\MSU}
\affiliation{\Duluth}
\affiliation{\Minnesota}
\affiliation{\Mississippi}
\affiliation{\NISER}
\affiliation{\Panjab}
\affiliation{\Pitt}
\affiliation{\QMU}
\affiliation{\SAlabama}
\affiliation{\Carolina}
\affiliation{\SMU}
\affiliation{\Sussex}
\affiliation{\Syracuse}
\affiliation{\Texas}
\affiliation{\Tufts}
\affiliation{\UCL}
\affiliation{\Virginia}
\affiliation{\WSU}
\affiliation{\WandM}
\affiliation{\Wisconsin}

\author{M.~A.~Acero}
\affiliation{\Atlantico}

\author{B.~Acharya}
\affiliation{\Mississippi}

\author{P.~Adamson}
\affiliation{\FNAL}









\author{N.~Anfimov}
\affiliation{\JINR}


\author{A.~Antoshkin}
\affiliation{\JINR}


\author{E.~Arrieta-Diaz}
\affiliation{\Magdalena}

\author{L.~Asquith}
\affiliation{\Sussex}


\author{A.~Aurisano}
\affiliation{\Cincinnati}


\author{A.~Back}
\affiliation{\Indiana}
\affiliation{\Iowa}



\author{N.~Balashov}
\affiliation{\JINR}

% ...

\collaboration{The NOvA Collaboration}
\noaffiliation

\date{\today}

\begin{abstract}
We report a search for neutrino oscillations to sterile neutrinos under a model with three active and one sterile neutrinos (3+1 model). This analysis uses the NOvA detectors exposed to the NuMI beam, running in neutrino mode. The data exposure, \num{13.6e20} protons on target, doubles that previously analyzed by NOvA, and the analysis is the first to use $\nu_{\mu}$ charged-current interactions in conjunction with neutral-current interactions. Neutrino samples in the Near and Far detectors are fitted simultaneously, enabling the search to be carried out over a $\Delta m^2_{41}$  range extending 2 (3) orders of magnitude above (below) \qty{1}{eV^2}. NOvA finds no evidence for active-to-sterile neutrino oscillations under the 3+1 model at \qty{90}{\percent} confidence level. New limits are reported in multiple regions of parameter space, excluding some regions currently allowed by IceCube at \qty{90}{\percent} confidence level. We additionally set the most stringent limits for anomalous $\nu_{\tau}$ appearance for $\Delta m^{2}_{41} \le \qty{3}{eV^2}$.
\end{abstract}


\maketitle


Neutrino mixing is a well established phenomenon, with numerous experiments reporting results that agree with a picture including three neutrino mass states ($\nu_{1}$, $\nu_{2}$, $\nu_{3}$) that mix to form three neutrino flavors ($\nu_{\mu}$, $\nu_{e}$, $\nu_{\tau}$)~\cite{Super-Kamiokande:1998kpq,SNO:2002tuh,KamLAND:2004mhv,K2K:2006yov,MINOS:2006foh,T2K:2011ypd,DayaBay:2012fng,RENO:2012mkc,DoubleChooz:2014kuw,OPERA:2015wbl,NOvA:2016kwd}. In this three-flavor (3F) framework, oscillations are governed by two mass-squared splittings, $\Delta m^{2}_{21}$ and $\Delta m^{2}_{32}$, where $\Delta m_{ji}^{2} \equiv m_{j}^{2} - m_{i}^{2}$, which correspond to the frequency of oscillation for a given neutrino energy ($E_{\nu}$) and path length ($L$), three mixing angles, $\theta_{12}$, $\theta_{13}$ and $\theta_{23}$, which drive the magnitude of oscillation, and a CP violating phase, $\delta_{\mathrm{CP}}$, which allows for differences between neutrino and antineutrino oscillations. Over the past two decades, a number of anomalous results have been reported in short baseline accelerator neutrino experiments~\cite{LSND:2001aii,MiniBooNE:2020pnu}, radiochemical experiments~\cite{Barinov:2022wfh,SAGE:1998fvr,GALLEX:1997lja}, and in the reactor neutrino sector~\cite{Huber:2011wv}. If these anomalies are interpreted as neutrino oscillations, they would require $\Delta m^2 \sim \qty{1}{eV^2} \gg \Delta m^2_{32}, \Delta m^2_{21}$, necessitating additional neutrino mass states be added to the model. Measurements of the width of the $Z$ boson at the LEP experiments indicate that any additional neutrinos with $m_{\nu} < m_{Z^{0}}/2$ must be sterile \cite{ALEPH:2005ab}, meaning that they do not interact via the weak force. The global picture of a mass splitting in the \qty{\sim1}{\eV^2} region is complicated by the presence of a number of null results \cite{KARMEN:2002zcm,MicroBooNE:2021tya,MINOS:2020iqj,Dydak:1983zq, Stockdale:1984cg, T2K:2019efw, SciBooNE:2011qyf, Super-Kamiokande:2014ndf, MINOS:2016viw, MINOS:2020iqj,  CCFRNuTeV:1998gjj, FERMILABE531:1986uhg, CHORUS:2007wlo, NOMAD:2001xxt,OPERA:2019kzo,IceCube:2017ivd}. 

NOvA can probe for active-to-sterile neutrino oscillations by searching for disappearance of neutral-current (NC) interactions, which provides a flavor-agnostic measurement of the active neutrino event rate. We can additionally search for active-to-sterile oscillations among $\nu_{\mu}$ charged-current (CC) interactions by testing for additional sources of disappearance when compared to 3F oscillations. The NOvA experiment consists of two functionally identical detectors placed \qty{14.6}{\milli\radian} off-axis of Fermilab's NuMI beam \cite{Adamson:2015dkw}. The NuMI beam is extracted over \qty{10}{\us} approximately every \qty{1.3}{\s} when \qty{120}{\GeV} protons strike a graphite target resulting in a secondary hadron beam. These hadrons are focused using two magnetic horns, and decay to neutrinos as they travel through a \qty{675}{m} helium-filled decay pipe. The off-axis placement of the detectors results in a narrow neutrino-energy distribution peaked around \qty{2}{\GeV}, with a width of \qty{0.4}{\GeV} and a sub-dominant high-energy tail.

The NOvA Near Detector (ND) is positioned at Fermilab in Batavia, Illinois, \qty{1}{\km} downstream of the NuMI target and \qty{100}{\m} underground, while the Far Detector (FD) is located \qty{810}{\km} from the target, at Ash River, Minnesota, on the surface with a \qty{3}{\m} water-equivalent overburden. The NOvA detectors are tracking calorimeters with the ND (FD) constructed of 192 (896) planes of highly reflective PVC cells measuring \qty{3.9}{\cm} x \qty{6.6}{\cm} with a length of \qty{3.9}{\m} (\qty{15.5}{\m}) \cite{Talaga:2016rlq}. The planes are alternately placed with horizontal and vertical cells to enable three-dimensional reconstruction, and are filled with a blend of mineral oil based liquid scintilator doped with \qty{5}{\percent} pseudocumene \cite{Mufson:2015kga}. The ND has additional planes of instrumented cells separated by steel plates at the rear of the detector (``muon catcher'') to range out muons. Light produced by charged particles traversing a cell is collected by a single loop of wavelength-shifting fiber which spans the length of the cell and is read out on both ends of the fiber at one end of the cell by one pixel of a \num{32} pixel avalanche photodiode (APD). Custom readout electronics are used to shape and digitize the signal, and any signal meeting a minimum pulse height requirement within a \qty{550}{\micro\s} window around the beam pulse is saved for offline analysis. The cosmic background at the FD is sampled by a \qty{10}{\Hz} minimum bias trigger~\cite{NOvA:2020dll}.

The simplest extension to 3F mixing is the 3+1 model~\cite{Caldwell:1993kn, Peltoniemi:1993ec, Bilenky:1998dt, Barger:2000ch, Goldman:2000sc}, which introduces seven new parameters: $\Delta m^{2}_{41}$, $\theta_{14}$, $\theta_{24}$, $\theta_{34}$, $\delta_{14}$, $\delta_{24}$, and $\delta_{34}$. Under this model, the active neutrino survival probability can be approximated as \cite{MINOS:2016viw}
\begin{equation}\label{eq:prob_nc}
    \begin{aligned}
    1 - P(\nu_{\mu} \rightarrow \nu_{s}) \approx 1 & - \cos^{4}\theta_{14}\cos^{2}\theta_{34}\sin^{2}2\theta_{24}\sin^{2}\Delta_{41} \\
    & - \sin^{2}\theta_{34}\sin^{2}2\theta_{23}\sin^{2}\Delta_{31} \\
    & + \frac{1}{2}\sin\delta_{24}\sin\theta_{24}\sin2\theta_{23}\sin\Delta_{31},
    \end{aligned}
\end{equation}
while the $\nu_{\mu}$ survival probability can be approximated as
\begin{equation}\label{eq:prob_cc_numu}
    \begin{aligned}
    P(\nu_{\mu} \rightarrow \nu_{\mu}) \approx 1 & - \sin^{2}2\theta_{24}\sin^{2}\Delta_{41} \\
    & + 2\sin^{2}2\theta_{23}\sin^{2}\theta_{24}\sin^{2}\Delta_{31} \\
    & - \sin^{2}2\theta_{23}\sin^{2}\Delta_{31},
    \end{aligned}
\end{equation}
where $\Delta_{ji} \equiv \frac{\Delta m_{ji}^{2} L}{4E_{\nu}}$. Exact oscillation probability calculations are used for the analysis.

\begin{figure}[!ht]
    \centering
    \begin{subfigure}[b]{\columnwidth}
      \includegraphics[width=\columnwidth]{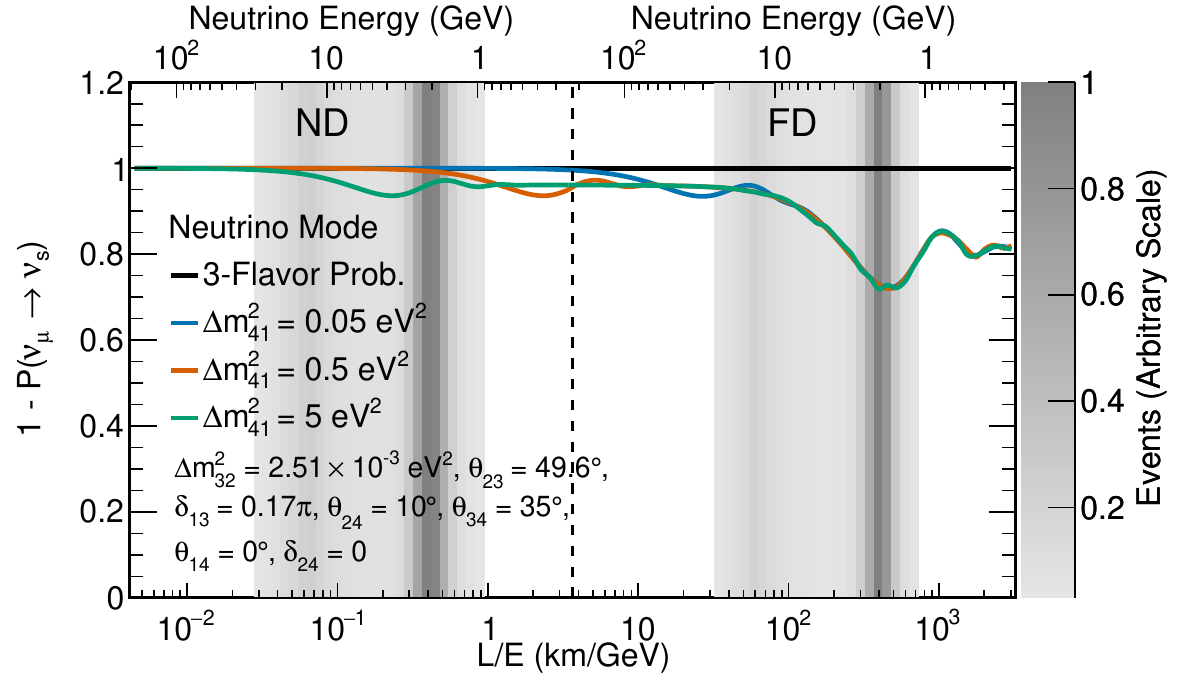}
      \caption{}
      \label{subfig:lovere_a}
    \end{subfigure}
        \begin{subfigure}[b]{\columnwidth}
      \includegraphics[width=\columnwidth]{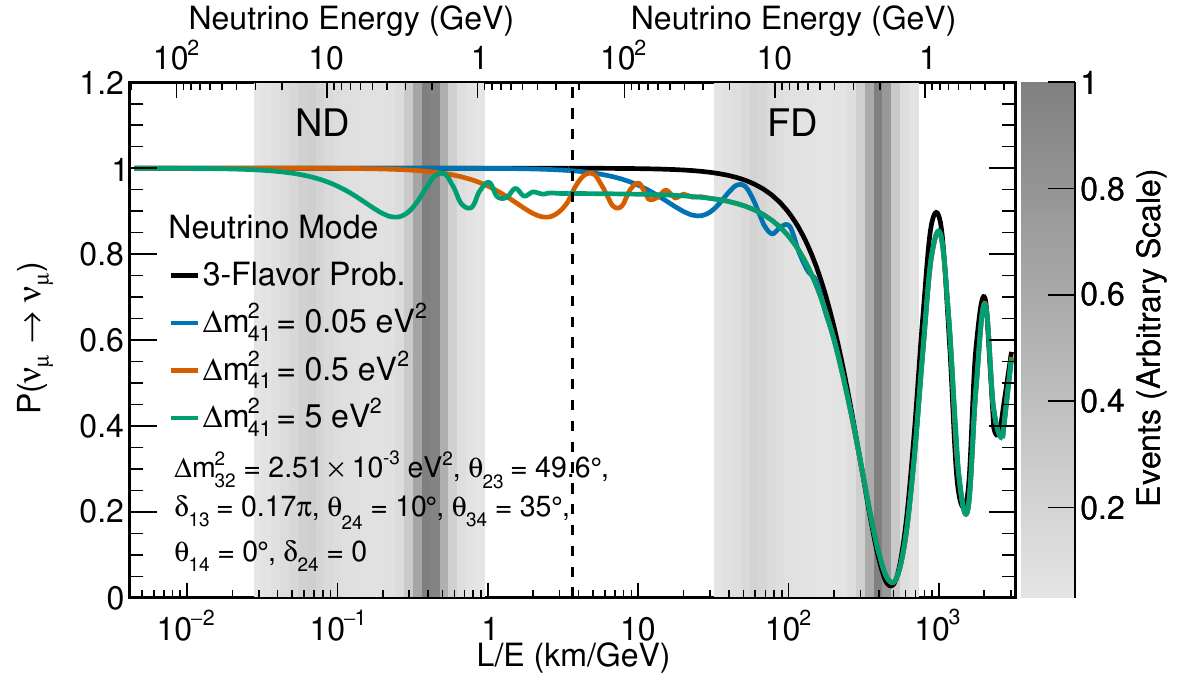}
      \caption{}
      \label{subfig:lovere_b}
    \end{subfigure}
    \caption{Oscillation probabilities for (a) active neutrino survival probability and (b) muon neutrino survival probability vs $E_{\nu}$ and $L/E_{\nu}$ with various model parameters. 3F survival probabilities are shown in black. Probabilities under the 3+1 model are shown with $\Delta m^2_{41}$ of \qty{0.05}{eV^2} (blue), \qty{0.5}{eV^2} (red), \qty{5}{eV^2} (green). As the value of $\Delta m^2_{41}$ increases, oscillations happen over shorter baselines, resulting in noticeable oscillations in the ND. The shaded regions approximately correspond to the fraction of neutrinos for each $E_{\nu}$ and $L/E_{\nu}$ in each detector.}
    \label{fig:lovere}
\end{figure}

Equations~\ref{eq:prob_nc} and~\ref{eq:prob_cc_numu} show that sterile neutrinos modify the magnitude of oscillations at the atmospheric frequency, $\Delta m^2_{31}$, and introduce a new sterile frequency driven by $\Delta m^2_{41}$. At the NOvA FD, the frequency of sterile oscillations at $\Delta m_{41}^{2} > \qty{0.05}{\eV^2}$ is too high for NOvA to resolve. These oscillations manifest as a downward normalization shift in the neutrino energy spectrum. At the shorter baseline of the NOvA ND, energy-dependent sterile oscillations arise when $\Delta m_{41}^{2} > \qty{0.5}{\eV^2}$ (Fig.~\ref{fig:lovere}). Fitting both detectors simultaneously extends the region of parameter space to which we are sensitive compared to previous analyses~\cite{NOvA:2017geg,NOvA:2021smv}. In the $\nu_{\mu}$ CC and NC channels considered, NOvA is sensitive to the atmospheric 3F oscillation parameters, $\theta_{23}$ and $\Delta m_{32}^{2}$, as well as to the sterile-related parameters $\theta_{24}$, $\theta_{34}$, $\Delta m_{41}^{2}$, and $\delta_{24}$, but not to $\theta_{14}$ as this analysis does not consider $\nu_{e}$ appearance.

This analysis uses data collected between February 2014 and March 2020, corresponding to \num{11.0e20} (\num{13.6e20}) protons on target (POT) for the ND (FD). Approximately \num{0.1e20} POT of ND NC selected events were removed from the sample for pre-analysis validation, meaning this sample corresponds to \num{10.9e20} POT.

The neutrino flux at the NOvA detectors is determined using a simulation of particle production and transport through the beamline using \textsc{Geant4} v9.2p03~\cite{Allison:2016lfl,Allison:2006ve,GEANT4:2002zbu}. The simulated flux is then corrected using the Package to Predict the Flux (PPFX) which modifies the prediction using external hadron-production data \cite{AliagaSoplin:2016shs}. Neutrino interactions are simulated within the NOvA detectors using GENIE 3.0.6~\cite{Andreopoulos:2009rq, Andreopoulos:2015wxa}. Additional information about the GENIE configuration used can be found in~\cite{NOvA:2021nfi}. The outgoing particles are propagated through the detector geometry using \textsc{Geant4} 10.4p02. Custom routines are used to simulate the capture and transport of scintillation light, as well as the response of the APDs.

The NC interaction candidates are characterised by hadronic activity resulting from the transfer of energy and momentum from the neutrino to the nucleus. The final state lepton is a neutrino and is not detected. All NC candidates are required to have a reconstructed vertex, at least \num{1} reconstructed particle, and must cross at least \num{3} contiguous planes.

In the ND, the reconstructed vertex is required to be contained within a volume with boundaries \qty{80}{\cm} from the top, bottom and sides of the detector, \qty{150}{\cm} from the front face, and \qty{260}{\cm} from the rear face excluding the muon catcher. All reconstructed particles are required to be contained within a volume excluding \qty{20}{\cm} from the top, bottom and sides of the detector, \qty{150}{\cm} from the front face, and \qty{50}{\cm} from the rear face excluding the muon catcher. The more stringent requirements on distance from the front and rear faces of the detector compared with the requirements on the other faces are selected to reject background candidates due to interactions in rock upstream of the detector and CC beam interactions, respectively.

For FD NC candidates, reconstructed particles are required to be fully contained within a fiducial volume with boundaries \qty{100}{\cm} from the top, bottom, and sides of the detector, and \qty{160}{\cm} from the upstream and downstream faces. The cosmic background interaction rate is significantly higher at the FD than the ND due to a shallower overburden. Accordingly, rather than placing explicit requirements on the vertex position, we use this information along with information about the reconstructed shower shapes and energy, number of hits, and the transverse momentum fraction to construct a Boosted Decision Tree focused on rejecting cosmic backgrounds. 

Signal events are selected using a Convolutional Neural Network (CNN) \cite{Aurisano:2016jvx,Psihas:2019ksa} which provides probability scores for different neutrino flavors based on energy depositions in the detector. Optimal requirements on the score are determined by using a figure of merit which considers the systematic and statistical uncertainties on the samples. In the FD, the requirements on CNN score and cosmic rejection score are jointly tuned. Any event passing the 3F $\nu_{\mu}$ CC or $\nu_{e}$ CC selection \cite{NOvA:2021nfi} is additionally removed from the sample of NC interaction candidates. 

The deposited energy of NC interaction candidates is estimated by taking a weighted sum of the hadronic and electromagnetic components of the calorimetric energy in the detector. An additional bias correction is applied as a function of total calorimetric energy. The overall NC energy resolution is 30\% \cite{Hausner:2022fli}. The event selection criteria and neutrino energy estimator used for the $\nu_{\mu}$ CC samples in the ND and FD are described in \cite{NOvA:2021nfi}.

We consider systematic uncertainties on the beam flux, neutrino interactions, and detector modeling \cite{NOvA:2021nfi}. For this analysis, we identified two sources of uncertainty that required custom handling.

Typically for oscillation analyses using NOvA's extrapolation technique~\cite{NOvA:2017geg, NOvA:2021smv, NOvA:2021nfi}, the ND is assumed to have no oscillations, so any differences between simulation and data can be attributed to mismodeling in the simulation. We then tune cross-section models in the ND simulation to the ND data, producing a new central value (CV) and suite of uncertainties. Because sterile neutrinos may induce oscillations in the ND, differences between data and simulation cannot be attributed to cross-section mismodeling, and so we use untuned simulation and uncertainties. Because the Meson Exchange Current (MEC) component of the simulation is the least well understood, we have developed shape and normalization uncertainties for this component based on the model spread of Val\`encia \cite{Nieves:2011pp}, SuSA \cite{Megias:2016fjk}, and GENIE empirical \cite{Katori:2013eoa} MEC models. 

Many NC neutrino candidates selected for this analysis are produced by kaon decays. Due to lack of available hadron-production data, prior analyses assigned the beam kaon component a \qty{30}{\percent} normalization uncertainty in addition to PPFX uncertainties. We instead constrain this uncertainty with samples not used in the analyses: a horn-off data sample, which allows us to probe hadron-production uncertainties without the complications of the focusing horns, and a sample of uncontained high-energy muon neutrinos, which gives us access to the focused kaon peak. We fit for the kaon component normalization marginalizing over potential sterile oscillations across the region of parameter space used in this analysis. This technique results in a \qty{10}{\percent} uncertainty on this component.

For each systematic uncertainty we randomly vary model parameters within their uncertainties to produce a new systematically-fluctuated ``universe,'' $u$. A covariance matrix is constructed,
\begin{equation}
    C_{i,j} = \frac{1}{U}\Sigma^{U}_{u=1}[N^{\mathrm{CV}}_{i}-N^{u}_{i}]\times[N^{\mathrm{CV}}_{j}-N^{u}_{j}],
\end{equation}
where $N_{i(j)}$ represents the number of events in the $i$th ($j$th) energy bin and $U$ is the total number of universes. The $C_{i,j}$ for each systematic uncertainty are summed, producing a final systematic covariance matrix.

We use two independent analysis techniques for this search. Analysis 1 employs a hybrid test statistic combining a Poisson log-likelihood treatment of statistical uncertainties with a Gaussian multivariate treatment of systematic uncertainties. A covariance matrix encoding only the systematic uncertainties is used to fit for optimal systematic weights, $s_{\alpha i}$, for each oscillation channel $\alpha$ and analysis bin $i$. This test statistic is expressed as
\begin{equation}
    \begin{aligned}
    \chi^{2} = 2 &\sum_{i} \left[ S_{i} - O_{i} + O_{i} \log \left( \frac{O_{i}}{S_{i}} \right) \right] \\
    + &\sum_{ij} \sum_{\alpha \beta} (s_{\alpha i} - 1) C_{\alpha i \beta j}^{-1} (s_{\beta j} - 1),
    \end{aligned}
\end{equation}
where $O_{i}$ are the observed data, $S_{i} = \sum_{\alpha} s_{\alpha i} p_{\alpha i}$ represents the systematic weights applied to the prediction, $p_{\alpha i}$, and $C_{\alpha i \beta j}$ is a covariance matrix encoding the systematic uncertainty in each oscillation channel ($\alpha$, $\beta$) and analysis bin ($i$, $j$) and their correlations.

Analysis 2 employs a traditional Gaussian multivariate formalism,
\begin{equation}
    \chi^{2} = \sum_{i} \sum_{j} (P_{i} - O_{i}) (C_{ij} + V_{ij})^{-1} (P_{j} - O_{j}),
\end{equation}
where $P_{i(j)}$ ($O_{i(j)}$) is the number of predicted (observed) events in bin $i(j)$. This analysis adds statistical uncertainties to the diagonal of the covariance matrix using the combined Neyman--Pearson formalism, $V_{ij} = \frac{3}{(1/O_{i}) + (2 / P_{i})} \delta_{ij}$, which yields a smaller bias in best fit model parameters than either the Neyman or Pearson construction~\cite{Ji:2019yca}.

\begin{figure}[!ht]
    \centering
      \includegraphics[width=\columnwidth]{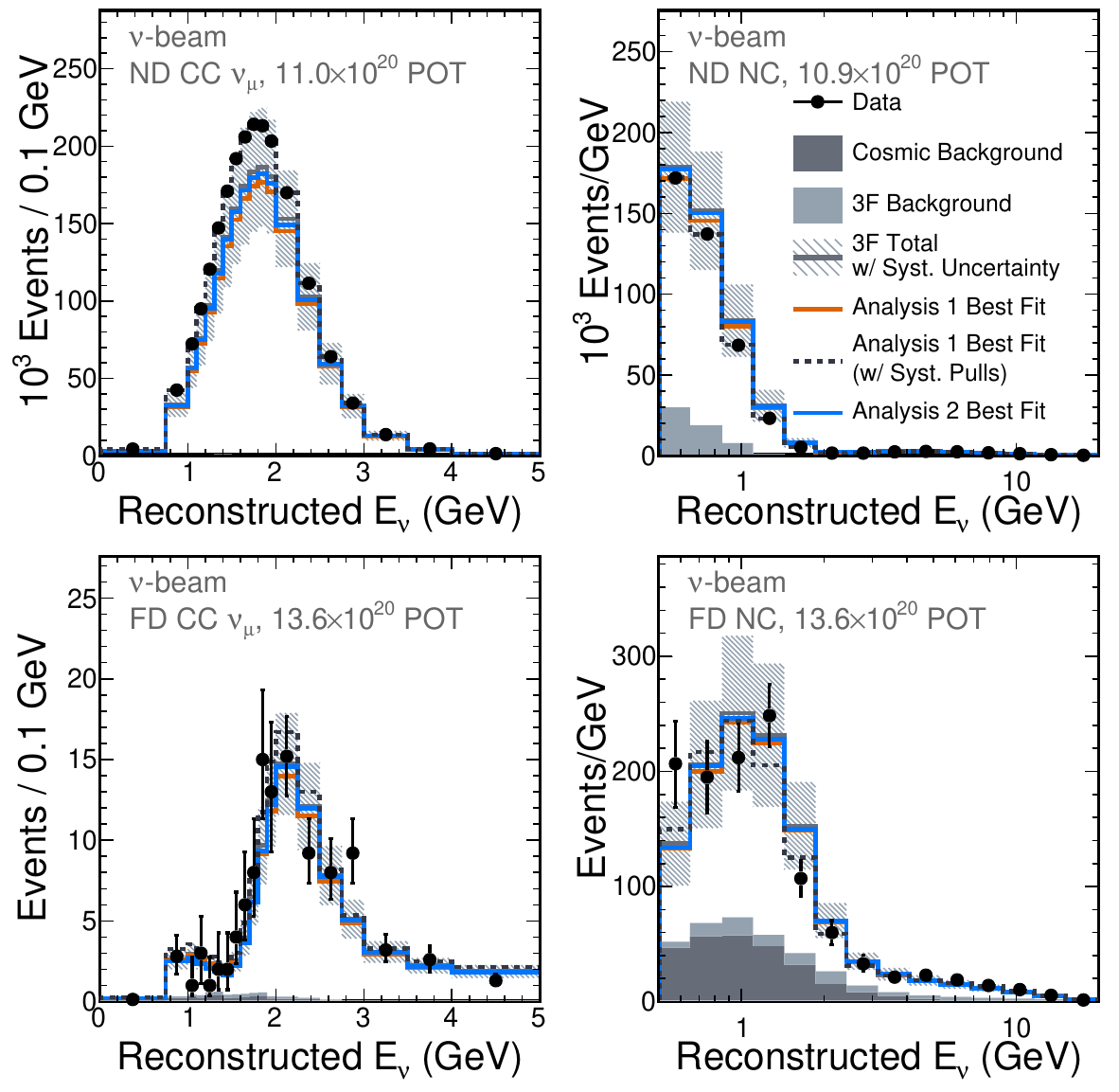}
    \caption{Reconstructed neutrino energy spectra for selected  $\nu_{\mu}$ CC (left) and NC (right) candidates in the ND (top) and FD (bottom). NOvA data shown as black points with 3F expectation shown as a gray histogram (mostly obscured) with shaded uncertainty bands. The backgrounds are shown as a stacked histogram in each panel, and are separated into cosmogenic data and simulated backgrounds under a 3F model with $\Delta m^2_{32} = 2.51 \times 10^{-3}~\mathrm{eV^2}$ and $\theta_{23} = \qty{49.6}{\degree}$. The best fits for Analysis 1 and Analysis 2 are shown in orange and blue, respectively. The dashed histogram shows the best fit for Analysis 1 with systematic pulls applied.}
    \label{fig:spectra}
\end{figure}

For both analyses the 3F atmospheric oscillation parameters $\theta_{23}$ and $\Delta m_{32}^{2}$ are varied in the fit, with a loose Gaussian constraint applied to $\Delta m_{32}^{2}$ to pin the fit to a $3+1$ flavor paradigm. This constraint is centered at $|$\num{2.51}$|$ $\pm$ \qty{0.15 e-3}{\eV^2} in both mass orderings, and is derived from a 2020 global fit to data \cite{Esteban:2020cvm} including atmospheric neutrino oscillations. The width of the constraint is conservatively set to double the $3 \sigma$ range. The sterile parameters $\Delta m_{41}^{2}$, $\theta_{24}$, $\theta_{34}$ and $\delta_{24}$ are also freely varied when fitting. The other sterile parameters are held fixed at \num{0} in the fit, due to constraints from solar and reactor experiments \cite{DayaBay:2022orm} and unitarity \cite{Parke:2015goa}.

We select \num{2826066} (\num{103109}) $\nu_{\mu}$ CC (NC) candidates from the ND data compared with the 3F prediction of \num{2450000} $\pm$ \num{530000} (\num{115000} $\pm$ \num{30000}). Additionally, we select \num{209}\footnote{Two events selected in \cite{NOvA:2021nfi} were not selected for this analysis due to an error in their hadronic energy reconstruction having been corrected by the time of this analysis.} (\num{469}) $\nu_{\mu}$ CC (NC) candidates from the FD data, compared with a 3F prediction of \num{200} $\pm$ \num{45} (\num{471} $\pm$ \num{116}) using pre-fit parameter values $\Delta m_{32}^{2} =$ \qty{2.51e-3}{eV^2} and $\theta_{23} =$ \qty{49.6}{\degree}~\cite{Esteban:2020cvm}. 

\begin{figure*}[!ht]
    \centering
    \begin{subfigure}[b]{0.66\columnwidth}
     \includegraphics[width=\columnwidth]{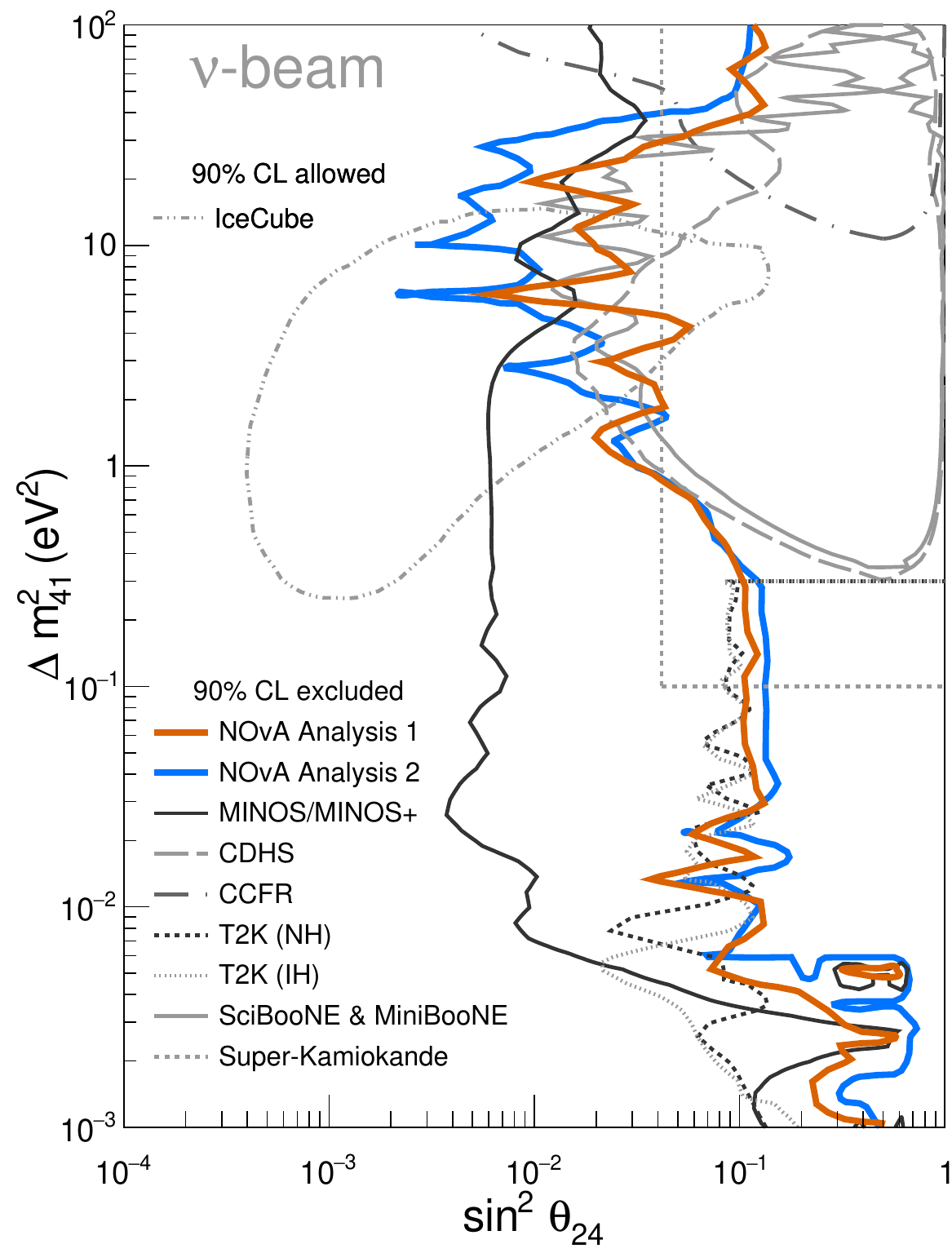}
      \caption{}
      \label{subfig:ssth24}
    \end{subfigure}
    \begin{subfigure}[b]{0.66\columnwidth}
      \includegraphics[width=\columnwidth]{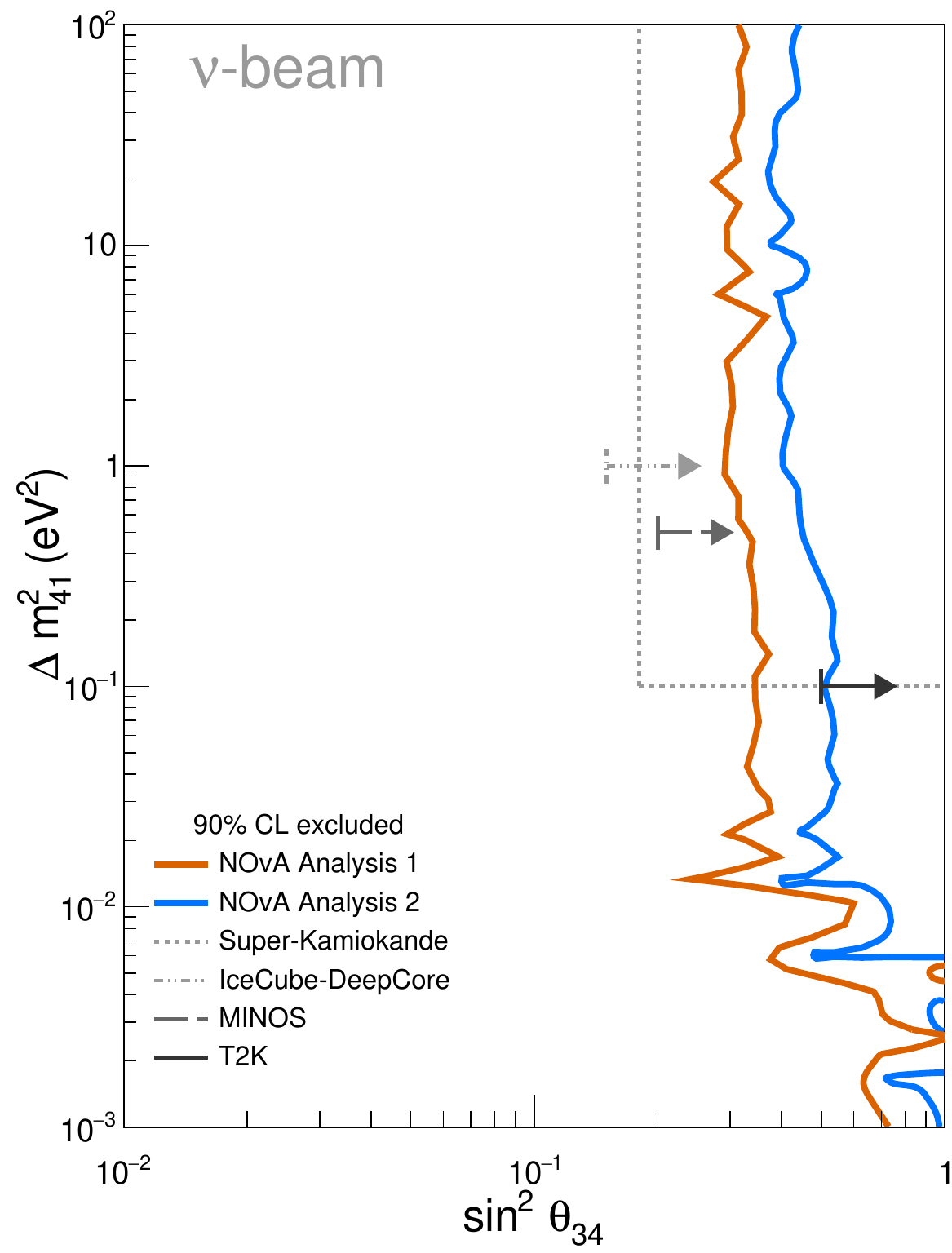}
      \caption{}
      \label{subfig:ssth34}
    \end{subfigure}
    \begin{subfigure}[b]{0.66\columnwidth}
        \includegraphics[width=\columnwidth]{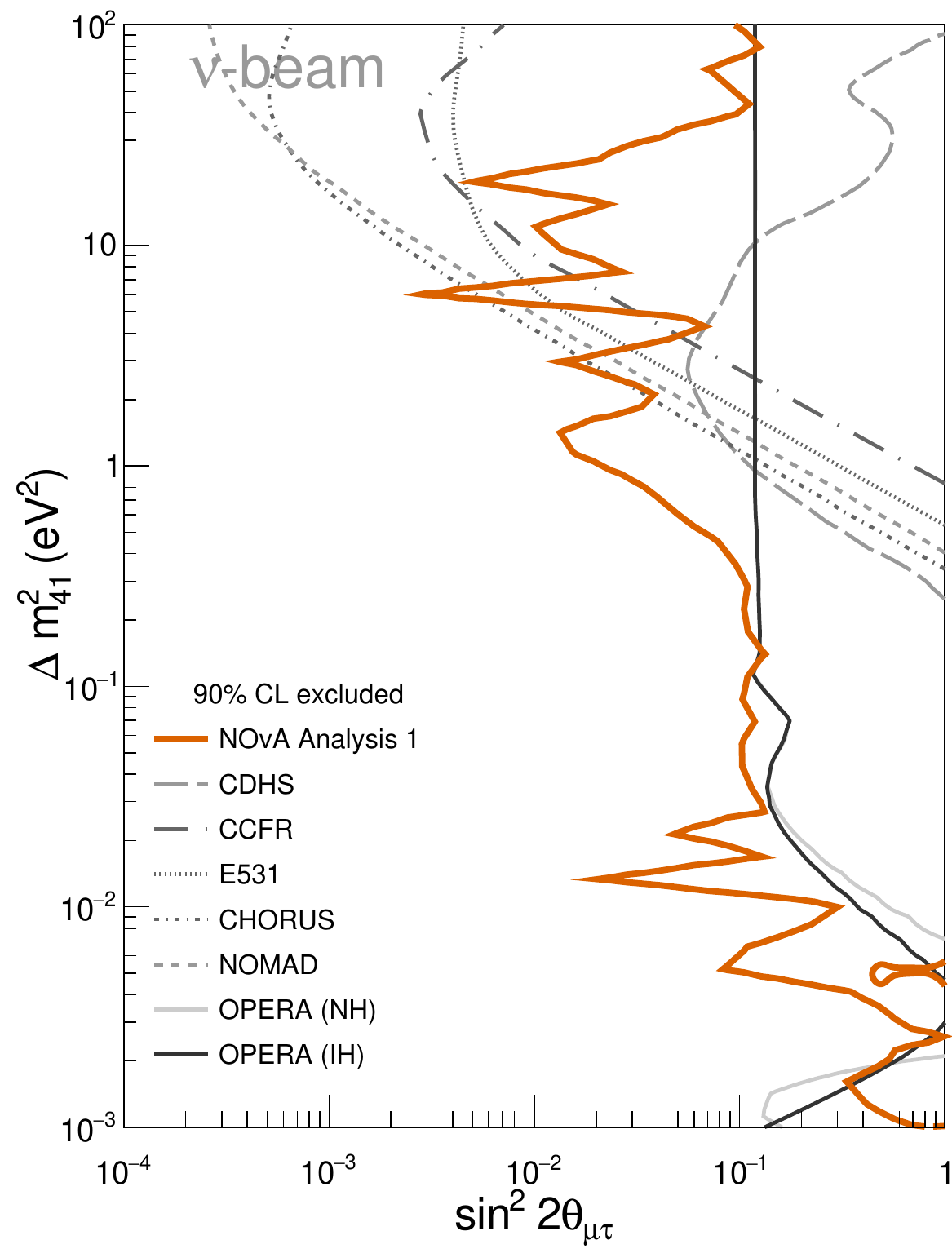}
        \caption{}
        \label{subfig:ssthmutau}
    \end{subfigure}
    \caption{NOvA's Feldman--Cousins corrected \qty{90}{\%} confidence limits in (a) $\Delta m_{41}^{2}-\sin^{2}\theta_{24}$ space, (b) $\Delta m_{41}^{2}-\sin^{2}\theta_{34}$ space, and (c) $\Delta m_{41}^{2}-\sin^{2}2\theta_{\mu \tau}$ space with allowed regions and exclusion contours from other experiments \cite{IceCube:2020phf, MINOS:2020iqj, Dydak:1983zq, Stockdale:1984cg, T2K:2019efw, SciBooNE:2011qyf, Super-Kamiokande:2014ndf, MINOS:2016viw, IceCube:2017ivd, CCFRNuTeV:1998gjj, FERMILABE531:1986uhg, CHORUS:2007wlo, NOMAD:2001xxt,OPERA:2019kzo}. Regions to the right of open contours are excluded. Closed contours for SciBooNE/MiniBooNE, CCFR, and CDHS in (a) also denote exclusion regions. For Super-Kamiokande, a single value of each mixing angle is reported for $\Delta m^2_{41} \ge$ \qty{0.1}{\eV^2} \cite{Super-Kamiokande:2014ndf}. Arrows in (b) represent a constraint on $\sin^2\theta_{34}$ at a single value of $\Delta m^2_{41}$ \cite{IceCube:2017ivd, MINOS:2016viw, T2K:2019efw}. OPERA NH/IH contours in (c) overlap at $\Delta m^2_{41} > 10^{-2} $ eV$^2$.}
    \label{fig:ssth24_ssth34}
\end{figure*}


The best fits from the two analyses agree well with the data and with each other (Fig.~\ref{fig:spectra}). The technique used by Analysis 1 allows us to present the best fit with systematic uncertainty pulls applied (dashed), resulting in improved agreement between data and simulation. This agreement indicates that any discrepancy between data and simulation can be accounted for by our systematic uncertainties.

We present \qty{90}{\percent} limits for both the $\Delta m^{2}_{41}-\sin^{2}\theta_{24}$ and $\Delta m^{2}_{41}-\sin^{2}\theta_{34}$ spaces (Figs. \ref{subfig:ssth24}, \ref{subfig:ssth34}). For each, a $\Delta\chi^2$ surface is constructed over a grid of the two parameters that define the space, with a fit performed for the remaining parameters at each point. To correct the critical $\chi^2$ values we use a hybrid Feldman--Cousins technique. Covariance matrix fitting techniques do not result in fitted pull terms for each systematic uncertainty. Accordingly, for systematic uncertainties we use a Highland--Cousins technique \cite{Cousins:1991qz}, where for each new universe a value of each systematic parameter is drawn from its a priori distribution. For oscillation parameters, we use the Profiled Feldman--Cousins technique \cite{NOvA:2022wnj}.

NOvA's $\Delta m^{2}_{41}-\sin^{2}\theta_{24}$ limits (Fig.~\ref{subfig:ssth24}) are competitive at high $\Delta m^2_{41}$, and exclude new regions of interest in the IceCube 90\% allowed region \cite{IceCube:2020phf}, around $\qty{6}{eV^2} < \Delta m^2_{41} < \qty{11}{eV^2}$. Analysis 1 excludes slightly more values of $\sin^{2}\theta_{24}$ at low $\Delta m^2_{41}$ and Analysis 2 excludes slightly more at high $\Delta m^2_{41}$. Sensitivity to $\sin^2{\theta_{24}}$ primarily comes from muon neutrino disappearance, which is independent of $\theta_{34}$ (Eq.~\ref{eq:prob_cc_numu}). For high values of $\Delta m^2_{41}$ sensitivity is driven by the ND data, meaning differences in the limits come from different handling of the systematic uncertainties. Sensitivity at low $\Delta m^2_{41}$ arises primarily from FD data, meaning that the weaker limit in Analysis 2 comes from under-coverage of the CNP statistical technique. 

Our $\Delta m^{2}_{41}-\sin^{2}\theta_{34}$ contours (Fig.~\ref{subfig:ssth34}) represent world-leading limits for $\Delta m^2_{41} < \qty{0.1}{eV^2}$. Sensitivity to $\sin^2\theta_{34}$ comes from our NC samples. For oscillations at the sterile frequency, oscillation probability $\propto\cos^2\theta_{34}\sin^2{\theta_{24}}$, resulting in reduced sensitivity to sterile oscillations in the ND (Eq.~\ref{eq:prob_nc}). For this space our sensitivity comes primarily from oscillations at the atmospheric frequency and therefore FD data, which are statistically limited and without strong dependence on $\Delta m^2_{41}$. In this space, Analysis 1 excludes slightly more parameter space than Analysis 2 across the full range considered. The differences in the contours in this space are attributed to the different statistical treatments of the two analyses.  

Finally, in Fig.~\ref{subfig:ssthmutau}, we present our results in terms of the effective mixing parameter, $\sin^2{2\theta_{\mu\tau}} = 4|U_{\mu4}|^2|U_{\tau4}|^2 = \sin^2\theta_{24}\sin^{2}\theta_{34}$,
which can be thought of as describing anomalous sterile-driven $\nu_{\tau}$ appearance. Because the analyses are consistent and the Feldman--Cousins procedure is resource intensive, we choose to present this contour using only Analysis 1. NOvA's ND is at a higher $L/E_{\nu}$ than other experiments with limits in this space, meaning that we are able to probe to lower values of $\Delta m^2_{41}$ resulting in the NOvA \qty{90}{\percent} limit being world-leading across large areas below $\Delta m^2_{41} = \qty{3}{eV^2}$. Notably, this limit excludes a new region of phase space around $\Delta m^2_{41} = \qty{1}{eV^2}$, the preferred region of $\Delta m^2_{41}$ for current anomalies.

In conclusion, an improved search for sterile neutrino oscillations under the 3+1 oscillation paradigm has been performed using NOvA data. We use two covariance matrix-based techniques that allow us to probe a wider range of $\Delta m^2_{41}$ values than previous NOvA analyses \cite{NOvA:2017geg, NOvA:2021smv}. Differences between the limits for the two analyses can be taken as an uncertainty due to analysis choices such as statistical treatment, systematic treatment, binning of the $\Delta\chi^2$ surface, and fitting technique. We find that the NOvA data are consistent with 3F oscillations at \qty{90}{\percent} confidence, and our limits agree with sensitivity studies performed using 3F oscillations. Our limits are the first presented in some regions of phase space, while excluding new regions of parameter space currently allowed by IceCube at \qty{90}{\%} confidence level. This work additionally sets the most stringent limits for anomalous $\nu_{\tau}$ appearance for $\Delta m^2_{41} \lesssim \qty{3}{eV^2}$, including the strongest limits around $\Delta m^2_{41} = \qty{1}{eV^2}$.

This document was prepared by the NOvA collaboration using the resources of the Fermi National Accelerator Laboratory (Fermilab), a U.S. Department of Energy, Office of Science, HEP User Facility. Fermilab is managed by Fermi Research Alliance, LLC (FRA), acting under Contract No. DE-AC02-07CH11359. This work was supported by the U.S. Department of Energy; the U.S. National Science Foundation; the Department of Science and Technology, India; the European Research Council; the MSMT CR, GA UK, Czech Republic; the RAS, MSHE, and RFBR, Russia; CNPq and FAPEG, Brazil; UKRI, STFC and the Royal Society, United Kingdom; and the state and University of Minnesota.  We are grateful for the contributions of the staffs of the University of Minnesota at the Ash River Laboratory, and of Fermilab.

\bibliographystyle{apsrev4-2}
\bibliography{bibliography}

\begin{thebibliography}{64}%
\makeatletter
\providecommand \@ifxundefined [1]{%
 \@ifx{#1\undefined}
}%
\providecommand \@ifnum [1]{%
 \ifnum #1\expandafter \@firstoftwo
 \else \expandafter \@secondoftwo
 \fi
}%
\providecommand \@ifx [1]{%
 \ifx #1\expandafter \@firstoftwo
 \else \expandafter \@secondoftwo
 \fi
}%
\providecommand \natexlab [1]{#1}%
\providecommand \enquote  [1]{``#1''}%
\providecommand \bibnamefont  [1]{#1}%
\providecommand \bibfnamefont [1]{#1}%
\providecommand \citenamefont [1]{#1}%
\providecommand \href@noop [0]{\@secondoftwo}%
\providecommand \href [0]{\begingroup \@sanitize@url \@href}%
\providecommand \@href[1]{\@@startlink{#1}\@@href}%
\providecommand \@@href[1]{\endgroup#1\@@endlink}%
\providecommand \@sanitize@url [0]{\catcode `\\12\catcode `\$12\catcode `\&12\catcode `\#12\catcode `\^12\catcode `\_12\catcode `\%12\relax}%
\providecommand \@@startlink[1]{}%
\providecommand \@@endlink[0]{}%
\providecommand \url  [0]{\begingroup\@sanitize@url \@url }%
\providecommand \@url [1]{\endgroup\@href {#1}{\urlprefix }}%
\providecommand \urlprefix  [0]{URL }%
\providecommand \Eprint [0]{\href }%
\providecommand \doibase [0]{https://doi.org/}%
\providecommand \selectlanguage [0]{\@gobble}%
\providecommand \bibinfo  [0]{\@secondoftwo}%
\providecommand \bibfield  [0]{\@secondoftwo}%
\providecommand \translation [1]{[#1]}%
\providecommand \BibitemOpen [0]{}%
\providecommand \bibitemStop [0]{}%
\providecommand \bibitemNoStop [0]{.\EOS\space}%
\providecommand \EOS [0]{\spacefactor3000\relax}%
\providecommand \BibitemShut  [1]{\csname bibitem#1\endcsname}%
\let\auto@bib@innerbib\@empty
\bibitem [{\citenamefont {Fukuda}\ \emph {et~al.}(1998)\citenamefont {Fukuda} \emph {et~al.}}]{Super-Kamiokande:1998kpq}%
  \BibitemOpen
  \bibfield  {author} {\bibinfo {author} {\bibfnamefont {Y.}~\bibnamefont {Fukuda}} \emph {et~al.} (\bibinfo {collaboration} {Super-Kamiokande}),\ }\href {https://doi.org/10.1103/PhysRevLett.81.1562} {\bibfield  {journal} {\bibinfo  {journal} {Phys. Rev. Lett.}\ }\textbf {\bibinfo {volume} {81}},\ \bibinfo {pages} {1562} (\bibinfo {year} {1998})},\ \Eprint {https://arxiv.org/abs/hep-ex/9807003} {arXiv:hep-ex/9807003} \BibitemShut {NoStop}%
\bibitem [{\citenamefont {Ahmad}\ \emph {et~al.}(2002)\citenamefont {Ahmad} \emph {et~al.}}]{SNO:2002tuh}%
  \BibitemOpen
  \bibfield  {author} {\bibinfo {author} {\bibfnamefont {Q.~R.}\ \bibnamefont {Ahmad}} \emph {et~al.} (\bibinfo {collaboration} {SNO}),\ }\href {https://doi.org/10.1103/PhysRevLett.89.011301} {\bibfield  {journal} {\bibinfo  {journal} {Phys. Rev. Lett.}\ }\textbf {\bibinfo {volume} {89}},\ \bibinfo {pages} {011301} (\bibinfo {year} {2002})},\ \Eprint {https://arxiv.org/abs/nucl-ex/0204008} {arXiv:nucl-ex/0204008} \BibitemShut {NoStop}%
\bibitem [{\citenamefont {Araki}\ \emph {et~al.}(2005)\citenamefont {Araki} \emph {et~al.}}]{KamLAND:2004mhv}%
  \BibitemOpen
  \bibfield  {author} {\bibinfo {author} {\bibfnamefont {T.}~\bibnamefont {Araki}} \emph {et~al.} (\bibinfo {collaboration} {KamLAND}),\ }\href {https://doi.org/10.1103/PhysRevLett.94.081801} {\bibfield  {journal} {\bibinfo  {journal} {Phys. Rev. Lett.}\ }\textbf {\bibinfo {volume} {94}},\ \bibinfo {pages} {081801} (\bibinfo {year} {2005})},\ \Eprint {https://arxiv.org/abs/hep-ex/0406035} {arXiv:hep-ex/0406035} \BibitemShut {NoStop}%
\bibitem [{\citenamefont {Ahn}\ \emph {et~al.}(2006)\citenamefont {Ahn} \emph {et~al.}}]{K2K:2006yov}%
  \BibitemOpen
  \bibfield  {author} {\bibinfo {author} {\bibfnamefont {M.~H.}\ \bibnamefont {Ahn}} \emph {et~al.} (\bibinfo {collaboration} {K2K}),\ }\href {https://doi.org/10.1103/PhysRevD.74.072003} {\bibfield  {journal} {\bibinfo  {journal} {Phys. Rev. D}\ }\textbf {\bibinfo {volume} {74}},\ \bibinfo {pages} {072003} (\bibinfo {year} {2006})},\ \Eprint {https://arxiv.org/abs/hep-ex/0606032} {arXiv:hep-ex/0606032} \BibitemShut {NoStop}%
\bibitem [{\citenamefont {Michael}\ \emph {et~al.}(2006)\citenamefont {Michael} \emph {et~al.}}]{MINOS:2006foh}%
  \BibitemOpen
  \bibfield  {author} {\bibinfo {author} {\bibfnamefont {D.~G.}\ \bibnamefont {Michael}} \emph {et~al.} (\bibinfo {collaboration} {MINOS}),\ }\href {https://doi.org/10.1103/PhysRevLett.97.191801} {\bibfield  {journal} {\bibinfo  {journal} {Phys. Rev. Lett.}\ }\textbf {\bibinfo {volume} {97}},\ \bibinfo {pages} {191801} (\bibinfo {year} {2006})},\ \Eprint {https://arxiv.org/abs/hep-ex/0607088} {arXiv:hep-ex/0607088} \BibitemShut {NoStop}%
\bibitem [{\citenamefont {Abe}\ \emph {et~al.}(2011)\citenamefont {Abe} \emph {et~al.}}]{T2K:2011ypd}%
  \BibitemOpen
  \bibfield  {author} {\bibinfo {author} {\bibfnamefont {K.}~\bibnamefont {Abe}} \emph {et~al.} (\bibinfo {collaboration} {T2K}),\ }\href {https://doi.org/10.1103/PhysRevLett.107.041801} {\bibfield  {journal} {\bibinfo  {journal} {Phys. Rev. Lett.}\ }\textbf {\bibinfo {volume} {107}},\ \bibinfo {pages} {041801} (\bibinfo {year} {2011})},\ \Eprint {https://arxiv.org/abs/1106.2822} {arXiv:1106.2822 [hep-ex]} \BibitemShut {NoStop}%
\bibitem [{\citenamefont {An}\ \emph {et~al.}(2012)\citenamefont {An} \emph {et~al.}}]{DayaBay:2012fng}%
  \BibitemOpen
  \bibfield  {author} {\bibinfo {author} {\bibfnamefont {F.~P.}\ \bibnamefont {An}} \emph {et~al.} (\bibinfo {collaboration} {Daya Bay}),\ }\href {https://doi.org/10.1103/PhysRevLett.108.171803} {\bibfield  {journal} {\bibinfo  {journal} {Phys. Rev. Lett.}\ }\textbf {\bibinfo {volume} {108}},\ \bibinfo {pages} {171803} (\bibinfo {year} {2012})},\ \Eprint {https://arxiv.org/abs/1203.1669} {arXiv:1203.1669 [hep-ex]} \BibitemShut {NoStop}%
\bibitem [{\citenamefont {Ahn}\ \emph {et~al.}(2012)\citenamefont {Ahn} \emph {et~al.}}]{RENO:2012mkc}%
  \BibitemOpen
  \bibfield  {author} {\bibinfo {author} {\bibfnamefont {J.~K.}\ \bibnamefont {Ahn}} \emph {et~al.} (\bibinfo {collaboration} {RENO}),\ }\href {https://doi.org/10.1103/PhysRevLett.108.191802} {\bibfield  {journal} {\bibinfo  {journal} {Phys. Rev. Lett.}\ }\textbf {\bibinfo {volume} {108}},\ \bibinfo {pages} {191802} (\bibinfo {year} {2012})},\ \Eprint {https://arxiv.org/abs/1204.0626} {arXiv:1204.0626 [hep-ex]} \BibitemShut {NoStop}%
\bibitem [{\citenamefont {Abe}\ \emph {et~al.}(2014)\citenamefont {Abe} \emph {et~al.}}]{DoubleChooz:2014kuw}%
  \BibitemOpen
  \bibfield  {author} {\bibinfo {author} {\bibfnamefont {Y.}~\bibnamefont {Abe}} \emph {et~al.} (\bibinfo {collaboration} {Double Chooz}),\ }\href {https://doi.org/10.1007/JHEP02(2015)074} {\bibfield  {journal} {\bibinfo  {journal} {JHEP}\ }\textbf {\bibinfo {volume} {10}},\ \bibinfo {pages} {086}},\ \bibinfo {note} {[Erratum: JHEP 02, 074 (2015)]},\ \Eprint {https://arxiv.org/abs/1406.7763} {arXiv:1406.7763 [hep-ex]} \BibitemShut {NoStop}%
\bibitem [{\citenamefont {Agafonova}\ \emph {et~al.}(2015)\citenamefont {Agafonova} \emph {et~al.}}]{OPERA:2015wbl}%
  \BibitemOpen
  \bibfield  {author} {\bibinfo {author} {\bibfnamefont {N.}~\bibnamefont {Agafonova}} \emph {et~al.} (\bibinfo {collaboration} {OPERA}),\ }\href {https://doi.org/10.1103/PhysRevLett.115.121802} {\bibfield  {journal} {\bibinfo  {journal} {Phys. Rev. Lett.}\ }\textbf {\bibinfo {volume} {115}},\ \bibinfo {pages} {121802} (\bibinfo {year} {2015})},\ \Eprint {https://arxiv.org/abs/1507.01417} {arXiv:1507.01417 [hep-ex]} \BibitemShut {NoStop}%
\bibitem [{\citenamefont {Adamson}\ \emph {et~al.}(2016{\natexlab{a}})\citenamefont {Adamson} \emph {et~al.}}]{NOvA:2016kwd}%
  \BibitemOpen
  \bibfield  {author} {\bibinfo {author} {\bibfnamefont {P.}~\bibnamefont {Adamson}} \emph {et~al.} (\bibinfo {collaboration} {NOvA}),\ }\href {https://doi.org/10.1103/PhysRevLett.116.151806} {\bibfield  {journal} {\bibinfo  {journal} {Phys. Rev. Lett.}\ }\textbf {\bibinfo {volume} {116}},\ \bibinfo {pages} {151806} (\bibinfo {year} {2016}{\natexlab{a}})},\ \Eprint {https://arxiv.org/abs/1601.05022} {arXiv:1601.05022 [hep-ex]} \BibitemShut {NoStop}%
\bibitem [{\citenamefont {Aguilar}\ \emph {et~al.}(2001)\citenamefont {Aguilar} \emph {et~al.}}]{LSND:2001aii}%
  \BibitemOpen
  \bibfield  {author} {\bibinfo {author} {\bibfnamefont {A.}~\bibnamefont {Aguilar}} \emph {et~al.} (\bibinfo {collaboration} {LSND}),\ }\href {https://doi.org/10.1103/PhysRevD.64.112007} {\bibfield  {journal} {\bibinfo  {journal} {Phys. Rev. D}\ }\textbf {\bibinfo {volume} {64}},\ \bibinfo {pages} {112007} (\bibinfo {year} {2001})},\ \Eprint {https://arxiv.org/abs/hep-ex/0104049} {arXiv:hep-ex/0104049} \BibitemShut {NoStop}%
\bibitem [{\citenamefont {Aguilar-Arevalo}\ \emph {et~al.}(2021)\citenamefont {Aguilar-Arevalo} \emph {et~al.}}]{MiniBooNE:2020pnu}%
  \BibitemOpen
  \bibfield  {author} {\bibinfo {author} {\bibfnamefont {A.~A.}\ \bibnamefont {Aguilar-Arevalo}} \emph {et~al.} (\bibinfo {collaboration} {MiniBooNE}),\ }\href {https://doi.org/10.1103/PhysRevD.103.052002} {\bibfield  {journal} {\bibinfo  {journal} {Phys. Rev. D}\ }\textbf {\bibinfo {volume} {103}},\ \bibinfo {pages} {052002} (\bibinfo {year} {2021})},\ \Eprint {https://arxiv.org/abs/2006.16883} {arXiv:2006.16883 [hep-ex]} \BibitemShut {NoStop}%
\bibitem [{\citenamefont {Barinov}\ \emph {et~al.}(2022)\citenamefont {Barinov} \emph {et~al.}}]{Barinov:2022wfh}%
  \BibitemOpen
  \bibfield  {author} {\bibinfo {author} {\bibfnamefont {V.~V.}\ \bibnamefont {Barinov}} \emph {et~al.},\ }\href {https://doi.org/10.1103/PhysRevC.105.065502} {\bibfield  {journal} {\bibinfo  {journal} {Phys. Rev. C}\ }\textbf {\bibinfo {volume} {105}},\ \bibinfo {pages} {065502} (\bibinfo {year} {2022})},\ \Eprint {https://arxiv.org/abs/2201.07364} {arXiv:2201.07364 [nucl-ex]} \BibitemShut {NoStop}%
\bibitem [{\citenamefont {Abdurashitov}\ \emph {et~al.}(1999)\citenamefont {Abdurashitov} \emph {et~al.}}]{SAGE:1998fvr}%
  \BibitemOpen
  \bibfield  {author} {\bibinfo {author} {\bibfnamefont {J.~N.}\ \bibnamefont {Abdurashitov}} \emph {et~al.} (\bibinfo {collaboration} {SAGE}),\ }\href {https://doi.org/10.1103/PhysRevC.59.2246} {\bibfield  {journal} {\bibinfo  {journal} {Phys. Rev. C}\ }\textbf {\bibinfo {volume} {59}},\ \bibinfo {pages} {2246} (\bibinfo {year} {1999})},\ \Eprint {https://arxiv.org/abs/hep-ph/9803418} {arXiv:hep-ph/9803418} \BibitemShut {NoStop}%
\bibitem [{\citenamefont {Hampel}\ \emph {et~al.}(1998)\citenamefont {Hampel} \emph {et~al.}}]{GALLEX:1997lja}%
  \BibitemOpen
  \bibfield  {author} {\bibinfo {author} {\bibfnamefont {W.}~\bibnamefont {Hampel}} \emph {et~al.} (\bibinfo {collaboration} {GALLEX}),\ }\href {https://doi.org/10.1016/S0370-2693(97)01562-1} {\bibfield  {journal} {\bibinfo  {journal} {Phys. Lett. B}\ }\textbf {\bibinfo {volume} {420}},\ \bibinfo {pages} {114} (\bibinfo {year} {1998})}\BibitemShut {NoStop}%
\bibitem [{\citenamefont {Huber}(2011)}]{Huber:2011wv}%
  \BibitemOpen
  \bibfield  {author} {\bibinfo {author} {\bibfnamefont {P.}~\bibnamefont {Huber}},\ }\href {https://doi.org/10.1103/PhysRevC.85.029901} {\bibfield  {journal} {\bibinfo  {journal} {Phys. Rev. C}\ }\textbf {\bibinfo {volume} {84}},\ \bibinfo {pages} {024617} (\bibinfo {year} {2011})},\ \bibinfo {note} {[Erratum: Phys.Rev.C 85, 029901 (2012)]},\ \Eprint {https://arxiv.org/abs/1106.0687} {arXiv:1106.0687 [hep-ph]} \BibitemShut {NoStop}%
\bibitem [{\citenamefont {Schael}\ \emph {et~al.}(2006)\citenamefont {Schael} \emph {et~al.}}]{ALEPH:2005ab}%
  \BibitemOpen
  \bibfield  {author} {\bibinfo {author} {\bibfnamefont {S.}~\bibnamefont {Schael}} \emph {et~al.} (\bibinfo {collaboration} {ALEPH, DELPHI, L3, OPAL, SLD, LEP Electroweak Working Group, SLD Electroweak Group, SLD Heavy Flavour Group}),\ }\href {https://doi.org/10.1016/j.physrep.2005.12.006} {\bibfield  {journal} {\bibinfo  {journal} {Phys. Rept.}\ }\textbf {\bibinfo {volume} {427}},\ \bibinfo {pages} {257} (\bibinfo {year} {2006})},\ \Eprint {https://arxiv.org/abs/hep-ex/0509008} {arXiv:hep-ex/0509008} \BibitemShut {NoStop}%
\bibitem [{\citenamefont {Armbruster}\ \emph {et~al.}(2002)\citenamefont {Armbruster} \emph {et~al.}}]{KARMEN:2002zcm}%
  \BibitemOpen
  \bibfield  {author} {\bibinfo {author} {\bibfnamefont {B.}~\bibnamefont {Armbruster}} \emph {et~al.} (\bibinfo {collaboration} {KARMEN}),\ }\href {https://doi.org/10.1103/PhysRevD.65.112001} {\bibfield  {journal} {\bibinfo  {journal} {Phys. Rev. D}\ }\textbf {\bibinfo {volume} {65}},\ \bibinfo {pages} {112001} (\bibinfo {year} {2002})},\ \Eprint {https://arxiv.org/abs/hep-ex/0203021} {arXiv:hep-ex/0203021} \BibitemShut {NoStop}%
\bibitem [{\citenamefont {Abratenko}\ \emph {et~al.}(2022)\citenamefont {Abratenko} \emph {et~al.}}]{MicroBooNE:2021tya}%
  \BibitemOpen
  \bibfield  {author} {\bibinfo {author} {\bibfnamefont {P.}~\bibnamefont {Abratenko}} \emph {et~al.} (\bibinfo {collaboration} {MicroBooNE}),\ }\href {https://doi.org/10.1103/PhysRevLett.128.241801} {\bibfield  {journal} {\bibinfo  {journal} {Phys. Rev. Lett.}\ }\textbf {\bibinfo {volume} {128}},\ \bibinfo {pages} {241801} (\bibinfo {year} {2022})},\ \Eprint {https://arxiv.org/abs/2110.14054} {arXiv:2110.14054 [hep-ex]} \BibitemShut {NoStop}%
\bibitem [{\citenamefont {Adamson}\ \emph {et~al.}(2020)\citenamefont {Adamson} \emph {et~al.}}]{MINOS:2020iqj}%
  \BibitemOpen
  \bibfield  {author} {\bibinfo {author} {\bibfnamefont {P.}~\bibnamefont {Adamson}} \emph {et~al.} (\bibinfo {collaboration} {MINOS+, Daya Bay}),\ }\href {https://doi.org/10.1103/PhysRevLett.125.071801} {\bibfield  {journal} {\bibinfo  {journal} {Phys. Rev. Lett.}\ }\textbf {\bibinfo {volume} {125}},\ \bibinfo {pages} {071801} (\bibinfo {year} {2020})},\ \Eprint {https://arxiv.org/abs/2002.00301} {arXiv:2002.00301 [hep-ex]} \BibitemShut {NoStop}%
\bibitem [{\citenamefont {Dydak}\ \emph {et~al.}(1984)\citenamefont {Dydak} \emph {et~al.}}]{Dydak:1983zq}%
  \BibitemOpen
  \bibfield  {author} {\bibinfo {author} {\bibfnamefont {F.}~\bibnamefont {Dydak}} \emph {et~al.},\ }\href {https://doi.org/10.1016/0370-2693(84)90688-9} {\bibfield  {journal} {\bibinfo  {journal} {Phys. Lett. B}\ }\textbf {\bibinfo {volume} {134}},\ \bibinfo {pages} {281} (\bibinfo {year} {1984})}\BibitemShut {NoStop}%
\bibitem [{\citenamefont {Stockdale}\ \emph {et~al.}(1984)\citenamefont {Stockdale} \emph {et~al.}}]{Stockdale:1984cg}%
  \BibitemOpen
  \bibfield  {author} {\bibinfo {author} {\bibfnamefont {I.~E.}\ \bibnamefont {Stockdale}} \emph {et~al.},\ }\href {https://doi.org/10.1103/PhysRevLett.52.1384} {\bibfield  {journal} {\bibinfo  {journal} {Phys. Rev. Lett.}\ }\textbf {\bibinfo {volume} {52}},\ \bibinfo {pages} {1384} (\bibinfo {year} {1984})}\BibitemShut {NoStop}%
\bibitem [{\citenamefont {Abe}\ \emph {et~al.}(2019)\citenamefont {Abe} \emph {et~al.}}]{T2K:2019efw}%
  \BibitemOpen
  \bibfield  {author} {\bibinfo {author} {\bibfnamefont {K.}~\bibnamefont {Abe}} \emph {et~al.} (\bibinfo {collaboration} {T2K}),\ }\href {https://doi.org/10.1103/PhysRevD.99.071103} {\bibfield  {journal} {\bibinfo  {journal} {Phys. Rev. D}\ }\textbf {\bibinfo {volume} {99}},\ \bibinfo {pages} {071103} (\bibinfo {year} {2019})},\ \Eprint {https://arxiv.org/abs/1902.06529} {arXiv:1902.06529 [hep-ex]} \BibitemShut {NoStop}%
\bibitem [{\citenamefont {Mahn}\ \emph {et~al.}(2012)\citenamefont {Mahn} \emph {et~al.}}]{SciBooNE:2011qyf}%
  \BibitemOpen
  \bibfield  {author} {\bibinfo {author} {\bibfnamefont {K.~B.~M.}\ \bibnamefont {Mahn}} \emph {et~al.} (\bibinfo {collaboration} {SciBooNE, MiniBooNE}),\ }\href {https://doi.org/10.1103/PhysRevD.85.032007} {\bibfield  {journal} {\bibinfo  {journal} {Phys. Rev. D}\ }\textbf {\bibinfo {volume} {85}},\ \bibinfo {pages} {032007} (\bibinfo {year} {2012})},\ \Eprint {https://arxiv.org/abs/1106.5685} {arXiv:1106.5685 [hep-ex]} \BibitemShut {NoStop}%
\bibitem [{\citenamefont {Abe}\ \emph {et~al.}(2015)\citenamefont {Abe} \emph {et~al.}}]{Super-Kamiokande:2014ndf}%
  \BibitemOpen
  \bibfield  {author} {\bibinfo {author} {\bibfnamefont {K.}~\bibnamefont {Abe}} \emph {et~al.} (\bibinfo {collaboration} {Super-Kamiokande}),\ }\href {https://doi.org/10.1103/PhysRevD.91.052019} {\bibfield  {journal} {\bibinfo  {journal} {Phys. Rev. D}\ }\textbf {\bibinfo {volume} {91}},\ \bibinfo {pages} {052019} (\bibinfo {year} {2015})},\ \Eprint {https://arxiv.org/abs/1410.2008} {arXiv:1410.2008 [hep-ex]} \BibitemShut {NoStop}%
\bibitem [{\citenamefont {Adamson}\ \emph {et~al.}(2016{\natexlab{b}})\citenamefont {Adamson} \emph {et~al.}}]{MINOS:2016viw}%
  \BibitemOpen
  \bibfield  {author} {\bibinfo {author} {\bibfnamefont {P.}~\bibnamefont {Adamson}} \emph {et~al.} (\bibinfo {collaboration} {MINOS}),\ }\href {https://doi.org/10.1103/PhysRevLett.117.151803} {\bibfield  {journal} {\bibinfo  {journal} {Phys. Rev. Lett.}\ }\textbf {\bibinfo {volume} {117}},\ \bibinfo {pages} {151803} (\bibinfo {year} {2016}{\natexlab{b}})},\ \Eprint {https://arxiv.org/abs/1607.01176} {arXiv:1607.01176 [hep-ex]} \BibitemShut {NoStop}%
\bibitem [{\citenamefont {Naples}\ \emph {et~al.}(1999)\citenamefont {Naples} \emph {et~al.}}]{CCFRNuTeV:1998gjj}%
  \BibitemOpen
  \bibfield  {author} {\bibinfo {author} {\bibfnamefont {D.}~\bibnamefont {Naples}} \emph {et~al.} (\bibinfo {collaboration} {CCFR/NuTeV}),\ }\href {https://doi.org/10.1103/PhysRevD.59.031101} {\bibfield  {journal} {\bibinfo  {journal} {Phys. Rev. D}\ }\textbf {\bibinfo {volume} {59}},\ \bibinfo {pages} {031101} (\bibinfo {year} {1999})},\ \Eprint {https://arxiv.org/abs/hep-ex/9809023} {arXiv:hep-ex/9809023} \BibitemShut {NoStop}%
\bibitem [{\citenamefont {Ushida}\ \emph {et~al.}(1986)\citenamefont {Ushida} \emph {et~al.}}]{FERMILABE531:1986uhg}%
  \BibitemOpen
  \bibfield  {author} {\bibinfo {author} {\bibfnamefont {N.}~\bibnamefont {Ushida}} \emph {et~al.} (\bibinfo {collaboration} {FERMILAB E531}),\ }\href {https://doi.org/10.1103/PhysRevLett.57.2897} {\bibfield  {journal} {\bibinfo  {journal} {Phys. Rev. Lett.}\ }\textbf {\bibinfo {volume} {57}},\ \bibinfo {pages} {2897} (\bibinfo {year} {1986})}\BibitemShut {NoStop}%
\bibitem [{\citenamefont {Eskut}\ \emph {et~al.}(2008)\citenamefont {Eskut} \emph {et~al.}}]{CHORUS:2007wlo}%
  \BibitemOpen
  \bibfield  {author} {\bibinfo {author} {\bibfnamefont {E.}~\bibnamefont {Eskut}} \emph {et~al.} (\bibinfo {collaboration} {CHORUS}),\ }\href {https://doi.org/10.1016/j.nuclphysb.2007.10.023} {\bibfield  {journal} {\bibinfo  {journal} {Nucl. Phys. B}\ }\textbf {\bibinfo {volume} {793}},\ \bibinfo {pages} {326} (\bibinfo {year} {2008})},\ \Eprint {https://arxiv.org/abs/0710.3361} {arXiv:0710.3361 [hep-ex]} \BibitemShut {NoStop}%
\bibitem [{\citenamefont {Astier}\ \emph {et~al.}(2001)\citenamefont {Astier} \emph {et~al.}}]{NOMAD:2001xxt}%
  \BibitemOpen
  \bibfield  {author} {\bibinfo {author} {\bibfnamefont {P.}~\bibnamefont {Astier}} \emph {et~al.} (\bibinfo {collaboration} {NOMAD}),\ }\href {https://doi.org/10.1016/S0550-3213(01)00339-X} {\bibfield  {journal} {\bibinfo  {journal} {Nucl. Phys. B}\ }\textbf {\bibinfo {volume} {611}},\ \bibinfo {pages} {3} (\bibinfo {year} {2001})},\ \Eprint {https://arxiv.org/abs/hep-ex/0106102} {arXiv:hep-ex/0106102} \BibitemShut {NoStop}%
\bibitem [{\citenamefont {Agafonova}\ \emph {et~al.}(2019)\citenamefont {Agafonova} \emph {et~al.}}]{OPERA:2019kzo}%
  \BibitemOpen
  \bibfield  {author} {\bibinfo {author} {\bibfnamefont {N.}~\bibnamefont {Agafonova}} \emph {et~al.} (\bibinfo {collaboration} {OPERA}),\ }\href {https://doi.org/10.1103/PhysRevD.100.051301} {\bibfield  {journal} {\bibinfo  {journal} {Phys. Rev. D}\ }\textbf {\bibinfo {volume} {100}},\ \bibinfo {pages} {051301} (\bibinfo {year} {2019})},\ \Eprint {https://arxiv.org/abs/1904.05686} {arXiv:1904.05686 [hep-ex]} \BibitemShut {NoStop}%
\bibitem [{\citenamefont {Aartsen}\ \emph {et~al.}(2017)\citenamefont {Aartsen} \emph {et~al.}}]{IceCube:2017ivd}%
  \BibitemOpen
  \bibfield  {author} {\bibinfo {author} {\bibfnamefont {M.~G.}\ \bibnamefont {Aartsen}} \emph {et~al.} (\bibinfo {collaboration} {IceCube}),\ }\href {https://doi.org/10.1103/PhysRevD.95.112002} {\bibfield  {journal} {\bibinfo  {journal} {Phys. Rev. D}\ }\textbf {\bibinfo {volume} {95}},\ \bibinfo {pages} {112002} (\bibinfo {year} {2017})},\ \Eprint {https://arxiv.org/abs/1702.05160} {arXiv:1702.05160 [hep-ex]} \BibitemShut {NoStop}%
\bibitem [{\citenamefont {Adamson}\ \emph {et~al.}(2016{\natexlab{c}})\citenamefont {Adamson} \emph {et~al.}}]{Adamson:2015dkw}%
  \BibitemOpen
  \bibfield  {author} {\bibinfo {author} {\bibfnamefont {P.}~\bibnamefont {Adamson}} \emph {et~al.},\ }\href {https://doi.org/10.1016/j.nima.2015.08.063} {\bibfield  {journal} {\bibinfo  {journal} {Nucl. Instrum. Meth. A}\ }\textbf {\bibinfo {volume} {806}},\ \bibinfo {pages} {279} (\bibinfo {year} {2016}{\natexlab{c}})},\ \Eprint {https://arxiv.org/abs/1507.06690} {arXiv:1507.06690 [physics.acc-ph]} \BibitemShut {NoStop}%
\bibitem [{\citenamefont {Talaga}\ \emph {et~al.}(2017)\citenamefont {Talaga}, \citenamefont {Grudzinski}, \citenamefont {Phan-Budd}, \citenamefont {Pla-Dalmau}, \citenamefont {Fagan}, \citenamefont {Grozis},\ and\ \citenamefont {Kephart}}]{Talaga:2016rlq}%
  \BibitemOpen
  \bibfield  {author} {\bibinfo {author} {\bibfnamefont {R.~L.}\ \bibnamefont {Talaga}}, \bibinfo {author} {\bibfnamefont {J.~J.}\ \bibnamefont {Grudzinski}}, \bibinfo {author} {\bibfnamefont {S.}~\bibnamefont {Phan-Budd}}, \bibinfo {author} {\bibfnamefont {A.}~\bibnamefont {Pla-Dalmau}}, \bibinfo {author} {\bibfnamefont {J.~E.}\ \bibnamefont {Fagan}}, \bibinfo {author} {\bibfnamefont {C.}~\bibnamefont {Grozis}},\ and\ \bibinfo {author} {\bibfnamefont {K.~M.}\ \bibnamefont {Kephart}},\ }\href {https://doi.org/10.1016/j.nima.2017.03.004} {\bibfield  {journal} {\bibinfo  {journal} {Nucl. Instrum. Meth. A}\ }\textbf {\bibinfo {volume} {861}},\ \bibinfo {pages} {77} (\bibinfo {year} {2017})},\ \Eprint {https://arxiv.org/abs/1601.00908} {arXiv:1601.00908 [physics.ins-det]} \BibitemShut {NoStop}%
\bibitem [{\citenamefont {Mufson}\ \emph {et~al.}(2015)\citenamefont {Mufson} \emph {et~al.}}]{Mufson:2015kga}%
  \BibitemOpen
  \bibfield  {author} {\bibinfo {author} {\bibfnamefont {S.}~\bibnamefont {Mufson}} \emph {et~al.},\ }\href {https://doi.org/10.1016/j.nima.2015.07.026} {\bibfield  {journal} {\bibinfo  {journal} {Nucl. Instrum. Meth. A}\ }\textbf {\bibinfo {volume} {799}},\ \bibinfo {pages} {1} (\bibinfo {year} {2015})},\ \Eprint {https://arxiv.org/abs/1504.04035} {arXiv:1504.04035 [physics.ins-det]} \BibitemShut {NoStop}%
\bibitem [{\citenamefont {Acero}\ \emph {et~al.}(2020)\citenamefont {Acero} \emph {et~al.}}]{NOvA:2020dll}%
  \BibitemOpen
  \bibfield  {author} {\bibinfo {author} {\bibfnamefont {M.~A.}\ \bibnamefont {Acero}} \emph {et~al.} (\bibinfo {collaboration} {NOvA}),\ }\href {https://doi.org/10.1088/1475-7516/2020/10/014} {\bibfield  {journal} {\bibinfo  {journal} {JCAP}\ }\textbf {\bibinfo {volume} {10}},\ \bibinfo {pages} {014}},\ \Eprint {https://arxiv.org/abs/2005.07155} {arXiv:2005.07155 [physics.ins-det]} \BibitemShut {NoStop}%
\bibitem [{\citenamefont {Caldwell}\ and\ \citenamefont {Mohapatra}(1993)}]{Caldwell:1993kn}%
  \BibitemOpen
  \bibfield  {author} {\bibinfo {author} {\bibfnamefont {D.~O.}\ \bibnamefont {Caldwell}}\ and\ \bibinfo {author} {\bibfnamefont {R.~N.}\ \bibnamefont {Mohapatra}},\ }\href {https://doi.org/10.1103/PhysRevD.48.3259} {\bibfield  {journal} {\bibinfo  {journal} {Phys. Rev. D}\ }\textbf {\bibinfo {volume} {48}},\ \bibinfo {pages} {3259} (\bibinfo {year} {1993})}\BibitemShut {NoStop}%
\bibitem [{\citenamefont {Peltoniemi}\ and\ \citenamefont {Valle}(1993)}]{Peltoniemi:1993ec}%
  \BibitemOpen
  \bibfield  {author} {\bibinfo {author} {\bibfnamefont {J.~T.}\ \bibnamefont {Peltoniemi}}\ and\ \bibinfo {author} {\bibfnamefont {J.~W.~F.}\ \bibnamefont {Valle}},\ }\href {https://doi.org/10.1016/0550-3213(93)90174-N} {\bibfield  {journal} {\bibinfo  {journal} {Nucl. Phys. B}\ }\textbf {\bibinfo {volume} {406}},\ \bibinfo {pages} {409} (\bibinfo {year} {1993})},\ \Eprint {https://arxiv.org/abs/hep-ph/9302316} {arXiv:hep-ph/9302316} \BibitemShut {NoStop}%
\bibitem [{\citenamefont {Bilenky}\ \emph {et~al.}(1999)\citenamefont {Bilenky}, \citenamefont {Giunti},\ and\ \citenamefont {Grimus}}]{Bilenky:1998dt}%
  \BibitemOpen
  \bibfield  {author} {\bibinfo {author} {\bibfnamefont {S.~M.}\ \bibnamefont {Bilenky}}, \bibinfo {author} {\bibfnamefont {C.}~\bibnamefont {Giunti}},\ and\ \bibinfo {author} {\bibfnamefont {W.}~\bibnamefont {Grimus}},\ }\href {https://doi.org/10.1016/S0146-6410(99)00092-7} {\bibfield  {journal} {\bibinfo  {journal} {Prog. Part. Nucl. Phys.}\ }\textbf {\bibinfo {volume} {43}},\ \bibinfo {pages} {1} (\bibinfo {year} {1999})},\ \Eprint {https://arxiv.org/abs/hep-ph/9812360} {arXiv:hep-ph/9812360} \BibitemShut {NoStop}%
\bibitem [{\citenamefont {Barger}\ \emph {et~al.}(2000)\citenamefont {Barger}, \citenamefont {Kayser}, \citenamefont {Learned}, \citenamefont {Weiler},\ and\ \citenamefont {Whisnant}}]{Barger:2000ch}%
  \BibitemOpen
  \bibfield  {author} {\bibinfo {author} {\bibfnamefont {V.~D.}\ \bibnamefont {Barger}}, \bibinfo {author} {\bibfnamefont {B.}~\bibnamefont {Kayser}}, \bibinfo {author} {\bibfnamefont {J.}~\bibnamefont {Learned}}, \bibinfo {author} {\bibfnamefont {T.~J.}\ \bibnamefont {Weiler}},\ and\ \bibinfo {author} {\bibfnamefont {K.}~\bibnamefont {Whisnant}},\ }\href {https://doi.org/10.1016/S0370-2693(00)00950-3} {\bibfield  {journal} {\bibinfo  {journal} {Phys. Lett. B}\ }\textbf {\bibinfo {volume} {489}},\ \bibinfo {pages} {345} (\bibinfo {year} {2000})},\ \Eprint {https://arxiv.org/abs/hep-ph/0008019} {arXiv:hep-ph/0008019} \BibitemShut {NoStop}%
\bibitem [{\citenamefont {Goldman}\ \emph {et~al.}(2000)\citenamefont {Goldman}, \citenamefont {Stephenson},\ and\ \citenamefont {McKellar}}]{Goldman:2000sc}%
  \BibitemOpen
  \bibfield  {author} {\bibinfo {author} {\bibfnamefont {J.~T.}\ \bibnamefont {Goldman}}, \bibinfo {author} {\bibfnamefont {G.~J.}\ \bibnamefont {Stephenson}, \bibfnamefont {Jr.}},\ and\ \bibinfo {author} {\bibfnamefont {B.~H.~J.}\ \bibnamefont {McKellar}},\ }\href {https://doi.org/10.1016/S0217-7323(00)00042-6} {\bibfield  {journal} {\bibinfo  {journal} {Mod. Phys. Lett. A}\ }\textbf {\bibinfo {volume} {15}},\ \bibinfo {pages} {439} (\bibinfo {year} {2000})},\ \Eprint {https://arxiv.org/abs/nucl-th/0002053} {arXiv:nucl-th/0002053} \BibitemShut {NoStop}%
\bibitem [{\citenamefont {Adamson}\ \emph {et~al.}(2017)\citenamefont {Adamson} \emph {et~al.}}]{NOvA:2017geg}%
  \BibitemOpen
  \bibfield  {author} {\bibinfo {author} {\bibfnamefont {P.}~\bibnamefont {Adamson}} \emph {et~al.} (\bibinfo {collaboration} {NOvA}),\ }\href {https://doi.org/10.1103/PhysRevD.96.072006} {\bibfield  {journal} {\bibinfo  {journal} {Phys. Rev. D}\ }\textbf {\bibinfo {volume} {96}},\ \bibinfo {pages} {072006} (\bibinfo {year} {2017})},\ \Eprint {https://arxiv.org/abs/1706.04592} {arXiv:1706.04592 [hep-ex]} \BibitemShut {NoStop}%
\bibitem [{\citenamefont {Acero}\ \emph {et~al.}(2021)\citenamefont {Acero} \emph {et~al.}}]{NOvA:2021smv}%
  \BibitemOpen
  \bibfield  {author} {\bibinfo {author} {\bibfnamefont {M.~A.}\ \bibnamefont {Acero}} \emph {et~al.} (\bibinfo {collaboration} {NOvA}),\ }\href {https://doi.org/10.1103/PhysRevLett.127.201801} {\bibfield  {journal} {\bibinfo  {journal} {Phys. Rev. Lett.}\ }\textbf {\bibinfo {volume} {127}},\ \bibinfo {pages} {201801} (\bibinfo {year} {2021})},\ \Eprint {https://arxiv.org/abs/2106.04673} {arXiv:2106.04673 [hep-ex]} \BibitemShut {NoStop}%
\bibitem [{\citenamefont {Allison}\ \emph {et~al.}(2016)\citenamefont {Allison} \emph {et~al.}}]{Allison:2016lfl}%
  \BibitemOpen
  \bibfield  {author} {\bibinfo {author} {\bibfnamefont {J.}~\bibnamefont {Allison}} \emph {et~al.},\ }\href {https://doi.org/10.1016/j.nima.2016.06.125} {\bibfield  {journal} {\bibinfo  {journal} {Nucl. Instrum. Meth. A}\ }\textbf {\bibinfo {volume} {835}},\ \bibinfo {pages} {186} (\bibinfo {year} {2016})}\BibitemShut {NoStop}%
\bibitem [{\citenamefont {Allison}\ \emph {et~al.}(2006)\citenamefont {Allison} \emph {et~al.}}]{Allison:2006ve}%
  \BibitemOpen
  \bibfield  {author} {\bibinfo {author} {\bibfnamefont {J.}~\bibnamefont {Allison}} \emph {et~al.},\ }\href {https://doi.org/10.1109/TNS.2006.869826} {\bibfield  {journal} {\bibinfo  {journal} {IEEE Trans. Nucl. Sci.}\ }\textbf {\bibinfo {volume} {53}},\ \bibinfo {pages} {270} (\bibinfo {year} {2006})}\BibitemShut {NoStop}%
\bibitem [{\citenamefont {Agostinelli}\ \emph {et~al.}(2003)\citenamefont {Agostinelli} \emph {et~al.}}]{GEANT4:2002zbu}%
  \BibitemOpen
  \bibfield  {author} {\bibinfo {author} {\bibfnamefont {S.}~\bibnamefont {Agostinelli}} \emph {et~al.} (\bibinfo {collaboration} {GEANT4}),\ }\href {https://doi.org/10.1016/S0168-9002(03)01368-8} {\bibfield  {journal} {\bibinfo  {journal} {Nucl. Instrum. Meth. A}\ }\textbf {\bibinfo {volume} {506}},\ \bibinfo {pages} {250} (\bibinfo {year} {2003})}\BibitemShut {NoStop}%
\bibitem [{\citenamefont {Aliaga~Soplin}(2016)}]{AliagaSoplin:2016shs}%
  \BibitemOpen
  \bibfield  {author} {\bibinfo {author} {\bibfnamefont {L.}~\bibnamefont {Aliaga~Soplin}},\ }\emph {\bibinfo {title} {{Neutrino Flux Prediction for the NuMI Beamline}}},\ \href {https://doi.org/10.2172/1250884} {Ph.D. thesis},\ \bibinfo  {school} {William-Mary Coll.} (\bibinfo {year} {2016})\BibitemShut {NoStop}%
\bibitem [{\citenamefont {Andreopoulos}\ \emph {et~al.}(2010)\citenamefont {Andreopoulos} \emph {et~al.}}]{Andreopoulos:2009rq}%
  \BibitemOpen
  \bibfield  {author} {\bibinfo {author} {\bibfnamefont {C.}~\bibnamefont {Andreopoulos}} \emph {et~al.},\ }\href {https://doi.org/10.1016/j.nima.2009.12.009} {\bibfield  {journal} {\bibinfo  {journal} {Nucl. Instrum. Meth. A}\ }\textbf {\bibinfo {volume} {614}},\ \bibinfo {pages} {87} (\bibinfo {year} {2010})},\ \Eprint {https://arxiv.org/abs/0905.2517} {arXiv:0905.2517 [hep-ph]} \BibitemShut {NoStop}%
\bibitem [{\citenamefont {Andreopoulos}\ \emph {et~al.}(2015)\citenamefont {Andreopoulos}, \citenamefont {Barry}, \citenamefont {Dytman}, \citenamefont {Gallagher}, \citenamefont {Golan}, \citenamefont {Hatcher}, \citenamefont {Perdue},\ and\ \citenamefont {Yarba}}]{Andreopoulos:2015wxa}%
  \BibitemOpen
  \bibfield  {author} {\bibinfo {author} {\bibfnamefont {C.}~\bibnamefont {Andreopoulos}}, \bibinfo {author} {\bibfnamefont {C.}~\bibnamefont {Barry}}, \bibinfo {author} {\bibfnamefont {S.}~\bibnamefont {Dytman}}, \bibinfo {author} {\bibfnamefont {H.}~\bibnamefont {Gallagher}}, \bibinfo {author} {\bibfnamefont {T.}~\bibnamefont {Golan}}, \bibinfo {author} {\bibfnamefont {R.}~\bibnamefont {Hatcher}}, \bibinfo {author} {\bibfnamefont {G.}~\bibnamefont {Perdue}},\ and\ \bibinfo {author} {\bibfnamefont {J.}~\bibnamefont {Yarba}},\ }\href@noop {} {\  (\bibinfo {year} {2015})},\ \Eprint {https://arxiv.org/abs/1510.05494} {arXiv:1510.05494 [hep-ph]} \BibitemShut {NoStop}%
\bibitem [{\citenamefont {Acero}\ \emph {et~al.}(2022{\natexlab{a}})\citenamefont {Acero} \emph {et~al.}}]{NOvA:2021nfi}%
  \BibitemOpen
  \bibfield  {author} {\bibinfo {author} {\bibfnamefont {M.~A.}\ \bibnamefont {Acero}} \emph {et~al.} (\bibinfo {collaboration} {NOvA}),\ }\href {https://doi.org/10.1103/PhysRevD.106.032004} {\bibfield  {journal} {\bibinfo  {journal} {Phys. Rev. D}\ }\textbf {\bibinfo {volume} {106}},\ \bibinfo {pages} {032004} (\bibinfo {year} {2022}{\natexlab{a}})},\ \Eprint {https://arxiv.org/abs/2108.08219} {arXiv:2108.08219 [hep-ex]} \BibitemShut {NoStop}%
\bibitem [{\citenamefont {Aurisano}\ \emph {et~al.}(2016)\citenamefont {Aurisano}, \citenamefont {Radovic}, \citenamefont {Rocco}, \citenamefont {Himmel}, \citenamefont {Messier}, \citenamefont {Niner}, \citenamefont {Pawloski}, \citenamefont {Psihas}, \citenamefont {Sousa},\ and\ \citenamefont {Vahle}}]{Aurisano:2016jvx}%
  \BibitemOpen
  \bibfield  {author} {\bibinfo {author} {\bibfnamefont {A.}~\bibnamefont {Aurisano}}, \bibinfo {author} {\bibfnamefont {A.}~\bibnamefont {Radovic}}, \bibinfo {author} {\bibfnamefont {D.}~\bibnamefont {Rocco}}, \bibinfo {author} {\bibfnamefont {A.}~\bibnamefont {Himmel}}, \bibinfo {author} {\bibfnamefont {M.~D.}\ \bibnamefont {Messier}}, \bibinfo {author} {\bibfnamefont {E.}~\bibnamefont {Niner}}, \bibinfo {author} {\bibfnamefont {G.}~\bibnamefont {Pawloski}}, \bibinfo {author} {\bibfnamefont {F.}~\bibnamefont {Psihas}}, \bibinfo {author} {\bibfnamefont {A.}~\bibnamefont {Sousa}},\ and\ \bibinfo {author} {\bibfnamefont {P.}~\bibnamefont {Vahle}},\ }\href {https://doi.org/10.1088/1748-0221/11/09/P09001} {\bibfield  {journal} {\bibinfo  {journal} {JINST}\ }\textbf {\bibinfo {volume} {11}}\bibfield  {number} {\bibinfo  {number} { (09)},\ \bibinfo {pages} {P09001}},\ }\Eprint {https://arxiv.org/abs/1604.01444} {arXiv:1604.01444 [hep-ex]} \BibitemShut {NoStop}%
\bibitem [{\citenamefont {Psihas}\ \emph {et~al.}(2019)\citenamefont {Psihas}, \citenamefont {Niner}, \citenamefont {Groh}, \citenamefont {Murphy}, \citenamefont {Aurisano}, \citenamefont {Himmel}, \citenamefont {Lang}, \citenamefont {Messier}, \citenamefont {Radovic},\ and\ \citenamefont {Sousa}}]{Psihas:2019ksa}%
  \BibitemOpen
  \bibfield  {author} {\bibinfo {author} {\bibfnamefont {F.}~\bibnamefont {Psihas}}, \bibinfo {author} {\bibfnamefont {E.}~\bibnamefont {Niner}}, \bibinfo {author} {\bibfnamefont {M.}~\bibnamefont {Groh}}, \bibinfo {author} {\bibfnamefont {R.}~\bibnamefont {Murphy}}, \bibinfo {author} {\bibfnamefont {A.}~\bibnamefont {Aurisano}}, \bibinfo {author} {\bibfnamefont {A.}~\bibnamefont {Himmel}}, \bibinfo {author} {\bibfnamefont {K.}~\bibnamefont {Lang}}, \bibinfo {author} {\bibfnamefont {M.~D.}\ \bibnamefont {Messier}}, \bibinfo {author} {\bibfnamefont {A.}~\bibnamefont {Radovic}},\ and\ \bibinfo {author} {\bibfnamefont {A.}~\bibnamefont {Sousa}},\ }\href {https://doi.org/10.1103/PhysRevD.100.073005} {\bibfield  {journal} {\bibinfo  {journal} {Phys. Rev. D}\ }\textbf {\bibinfo {volume} {100}},\ \bibinfo {pages} {073005} (\bibinfo {year} {2019})},\ \Eprint {https://arxiv.org/abs/1906.00713} {arXiv:1906.00713 [physics.ins-det]} \BibitemShut {NoStop}%
\bibitem [{\citenamefont {Hausner}(2022)}]{Hausner:2022fli}%
  \BibitemOpen
  \bibfield  {author} {\bibinfo {author} {\bibfnamefont {H.~R.}\ \bibnamefont {Hausner}},\ }\emph {\bibinfo {title} {{Sterile neutrino search with the NOvA detectors}}},\ \href@noop {} {Ph.D. thesis},\ \bibinfo  {school} {U. Wisconsin, Madison} (\bibinfo {year} {2022})\BibitemShut {NoStop}%
\bibitem [{\citenamefont {Nieves}\ \emph {et~al.}(2011)\citenamefont {Nieves}, \citenamefont {Ruiz~Simo},\ and\ \citenamefont {Vicente~Vacas}}]{Nieves:2011pp}%
  \BibitemOpen
  \bibfield  {author} {\bibinfo {author} {\bibfnamefont {J.}~\bibnamefont {Nieves}}, \bibinfo {author} {\bibfnamefont {I.}~\bibnamefont {Ruiz~Simo}},\ and\ \bibinfo {author} {\bibfnamefont {M.~J.}\ \bibnamefont {Vicente~Vacas}},\ }\href {https://doi.org/10.1103/PhysRevC.83.045501} {\bibfield  {journal} {\bibinfo  {journal} {Phys. Rev. C}\ }\textbf {\bibinfo {volume} {83}},\ \bibinfo {pages} {045501} (\bibinfo {year} {2011})},\ \Eprint {https://arxiv.org/abs/1102.2777} {arXiv:1102.2777 [hep-ph]} \BibitemShut {NoStop}%
\bibitem [{\citenamefont {Megias}\ \emph {et~al.}(2016)\citenamefont {Megias}, \citenamefont {Amaro}, \citenamefont {Barbaro}, \citenamefont {Caballero}, \citenamefont {Donnelly},\ and\ \citenamefont {Ruiz~Simo}}]{Megias:2016fjk}%
  \BibitemOpen
  \bibfield  {author} {\bibinfo {author} {\bibfnamefont {G.~D.}\ \bibnamefont {Megias}}, \bibinfo {author} {\bibfnamefont {J.~E.}\ \bibnamefont {Amaro}}, \bibinfo {author} {\bibfnamefont {M.~B.}\ \bibnamefont {Barbaro}}, \bibinfo {author} {\bibfnamefont {J.~A.}\ \bibnamefont {Caballero}}, \bibinfo {author} {\bibfnamefont {T.~W.}\ \bibnamefont {Donnelly}},\ and\ \bibinfo {author} {\bibfnamefont {I.}~\bibnamefont {Ruiz~Simo}},\ }\href {https://doi.org/10.1103/PhysRevD.94.093004} {\bibfield  {journal} {\bibinfo  {journal} {Phys. Rev. D}\ }\textbf {\bibinfo {volume} {94}},\ \bibinfo {pages} {093004} (\bibinfo {year} {2016})},\ \Eprint {https://arxiv.org/abs/1607.08565} {arXiv:1607.08565 [nucl-th]} \BibitemShut {NoStop}%
\bibitem [{\citenamefont {Katori}(2015)}]{Katori:2013eoa}%
  \BibitemOpen
  \bibfield  {author} {\bibinfo {author} {\bibfnamefont {T.}~\bibnamefont {Katori}},\ }\href {https://doi.org/10.1063/1.4919465} {\bibfield  {journal} {\bibinfo  {journal} {AIP Conf. Proc.}\ }\textbf {\bibinfo {volume} {1663}},\ \bibinfo {pages} {030001} (\bibinfo {year} {2015})},\ \Eprint {https://arxiv.org/abs/1304.6014} {arXiv:1304.6014 [nucl-th]} \BibitemShut {NoStop}%
\bibitem [{\citenamefont {Ji}\ \emph {et~al.}(2020)\citenamefont {Ji}, \citenamefont {Gu}, \citenamefont {Qian}, \citenamefont {Wei},\ and\ \citenamefont {Zhang}}]{Ji:2019yca}%
  \BibitemOpen
  \bibfield  {author} {\bibinfo {author} {\bibfnamefont {X.}~\bibnamefont {Ji}}, \bibinfo {author} {\bibfnamefont {W.}~\bibnamefont {Gu}}, \bibinfo {author} {\bibfnamefont {X.}~\bibnamefont {Qian}}, \bibinfo {author} {\bibfnamefont {H.}~\bibnamefont {Wei}},\ and\ \bibinfo {author} {\bibfnamefont {C.}~\bibnamefont {Zhang}},\ }\href {https://doi.org/10.1016/j.nima.2020.163677} {\bibfield  {journal} {\bibinfo  {journal} {Nucl. Instrum. Meth. A}\ }\textbf {\bibinfo {volume} {961}},\ \bibinfo {pages} {163677} (\bibinfo {year} {2020})},\ \Eprint {https://arxiv.org/abs/1903.07185} {arXiv:1903.07185 [physics.data-an]} \BibitemShut {NoStop}%
\bibitem [{\citenamefont {Esteban}\ \emph {et~al.}(2020)\citenamefont {Esteban}, \citenamefont {Gonzalez-Garcia}, \citenamefont {Maltoni}, \citenamefont {Schwetz},\ and\ \citenamefont {Zhou}}]{Esteban:2020cvm}%
  \BibitemOpen
  \bibfield  {author} {\bibinfo {author} {\bibfnamefont {I.}~\bibnamefont {Esteban}}, \bibinfo {author} {\bibfnamefont {M.~C.}\ \bibnamefont {Gonzalez-Garcia}}, \bibinfo {author} {\bibfnamefont {M.}~\bibnamefont {Maltoni}}, \bibinfo {author} {\bibfnamefont {T.}~\bibnamefont {Schwetz}},\ and\ \bibinfo {author} {\bibfnamefont {A.}~\bibnamefont {Zhou}},\ }\href {https://doi.org/10.1007/JHEP09(2020)178} {\bibfield  {journal} {\bibinfo  {journal} {JHEP}\ }\textbf {\bibinfo {volume} {09}},\ \bibinfo {pages} {178}},\ \Eprint {https://arxiv.org/abs/2007.14792} {arXiv:2007.14792 [hep-ph]} \BibitemShut {NoStop}%
\bibitem [{\citenamefont {An}\ \emph {et~al.}(2023)\citenamefont {An} \emph {et~al.}}]{DayaBay:2022orm}%
  \BibitemOpen
  \bibfield  {author} {\bibinfo {author} {\bibfnamefont {F.~P.}\ \bibnamefont {An}} \emph {et~al.} (\bibinfo {collaboration} {Daya Bay}),\ }\href {https://doi.org/10.1103/PhysRevLett.130.161802} {\bibfield  {journal} {\bibinfo  {journal} {Phys. Rev. Lett.}\ }\textbf {\bibinfo {volume} {130}},\ \bibinfo {pages} {161802} (\bibinfo {year} {2023})},\ \Eprint {https://arxiv.org/abs/2211.14988} {arXiv:2211.14988 [hep-ex]} \BibitemShut {NoStop}%
\bibitem [{\citenamefont {Parke}\ and\ \citenamefont {Ross-Lonergan}(2016)}]{Parke:2015goa}%
  \BibitemOpen
  \bibfield  {author} {\bibinfo {author} {\bibfnamefont {S.}~\bibnamefont {Parke}}\ and\ \bibinfo {author} {\bibfnamefont {M.}~\bibnamefont {Ross-Lonergan}},\ }\href {https://doi.org/10.1103/PhysRevD.93.113009} {\bibfield  {journal} {\bibinfo  {journal} {Phys. Rev. D}\ }\textbf {\bibinfo {volume} {93}},\ \bibinfo {pages} {113009} (\bibinfo {year} {2016})},\ \Eprint {https://arxiv.org/abs/1508.05095} {arXiv:1508.05095 [hep-ph]} \BibitemShut {NoStop}%
\bibitem [{\citenamefont {Aartsen}\ \emph {et~al.}(2020)\citenamefont {Aartsen} \emph {et~al.}}]{IceCube:2020phf}%
  \BibitemOpen
  \bibfield  {author} {\bibinfo {author} {\bibfnamefont {M.~G.}\ \bibnamefont {Aartsen}} \emph {et~al.} (\bibinfo {collaboration} {IceCube}),\ }\href {https://doi.org/10.1103/PhysRevLett.125.141801} {\bibfield  {journal} {\bibinfo  {journal} {Phys. Rev. Lett.}\ }\textbf {\bibinfo {volume} {125}},\ \bibinfo {pages} {141801} (\bibinfo {year} {2020})},\ \Eprint {https://arxiv.org/abs/2005.12942} {arXiv:2005.12942 [hep-ex]} \BibitemShut {NoStop}%
\bibitem [{\citenamefont {Cousins}\ and\ \citenamefont {Highland}(1992)}]{Cousins:1991qz}%
  \BibitemOpen
  \bibfield  {author} {\bibinfo {author} {\bibfnamefont {R.~D.}\ \bibnamefont {Cousins}}\ and\ \bibinfo {author} {\bibfnamefont {V.~L.}\ \bibnamefont {Highland}},\ }\href {https://doi.org/10.1016/0168-9002(92)90794-5} {\bibfield  {journal} {\bibinfo  {journal} {Nucl. Instrum. Meth. A}\ }\textbf {\bibinfo {volume} {320}},\ \bibinfo {pages} {331} (\bibinfo {year} {1992})}\BibitemShut {NoStop}%
\bibitem [{\citenamefont {Acero}\ \emph {et~al.}(2022{\natexlab{b}})\citenamefont {Acero} \emph {et~al.}}]{NOvA:2022wnj}%
  \BibitemOpen
  \bibfield  {author} {\bibinfo {author} {\bibfnamefont {M.~A.}\ \bibnamefont {Acero}} \emph {et~al.} (\bibinfo {collaboration} {NOvA}),\ }\href@noop {} {\  (\bibinfo {year} {2022}{\natexlab{b}})},\ \Eprint {https://arxiv.org/abs/2207.14353} {arXiv:2207.14353 [hep-ex]} \BibitemShut {NoStop}%
\end{thebibliography}%

\end{document}